\documentclass[conference]{IEEEtran}

\newcommand{\ifanonymous}[2]{%
  \ifcsname anonymous\endcsname%
    #1\ignorespaces
  \else%
    #2\ignorespaces
  \fi%
}

\makeatletter
\def\input@path{{./fig/}}
\makeatother

\usepackage{cite}
\usepackage[hyphens]{url}
\usepackage{subcaption}
\usepackage[english]{babel}
\usepackage{dashrule}
\usepackage{amsfonts}
\usepackage{amsthm}
\usepackage{amsmath}
\usepackage{amssymb}
\usepackage{enumitem}
\usepackage[T1]{fontenc}
\usepackage[utf8]{inputenc}
\usepackage{standalone}
\usepackage{multicol}
\usepackage{xcolor}
\usepackage{relsize}
\usepackage{listings}
\usepackage[switch]{lineno}
\usepackage{tikz}
\usepackage{tabularx}

\graphicspath{{fig/}}

\newcommand{\propSet}{\mathcal{P}}

\DeclareMathAlphabet{\mathcal}{OMS}{cmsy}{m}{n}
\SetMathAlphabet{\mathcal}{bold}{OMS}{cmsy}{b}{n}

\usepackage{customalgo}

\DeclareMathSymbol{\Omega}{\mathord}{operators}{"0A}

\let\emptyset\relax
\let\emptyset\varnothing

\newtheoremstyle{mystyle}
  {}
  {}
  {\itshape}
  {}
  {\bfseries}
  {.}
  { }
  {\thmname{#1}\thmnumber{ #2}\thmnote{ (#3)}} 

\theoremstyle{mystyle}
\newtheorem{mydef}{Definition}
\newtheorem{theorem}{Theorem}

\newtheorem{lemma}{Lemma}
\newtheorem{corollary}{Corollary}

\title{Linearizable State Machine Replication of State-Based CRDTs without Logs}

\ifanonymous{
  \author{Anonymous Authors}{ }{}{}{}
}{
\author{
  \IEEEauthorblockN{Jan Skrzypczak}
  \IEEEauthorblockA{\textit{Zuse Institute Berlin}\\
  Berlin, Germany \\
  skrzypczak@zib.de
  }
  \and
  \IEEEauthorblockN{Florian Schintke}
  \IEEEauthorblockA{\textit{Zuse Institute Berlin}\\
  Berlin, Germany \\
  schintke@zib.de
  }
  \and
  \IEEEauthorblockN{Thorsten Sch\"utt}
  \IEEEauthorblockA{\textit{Zuse Institute Berlin}\\
  Berlin, Germany \\
  schuett@zib.de
  }
}
}

\newcommand{\figref}[1]{\figurename~\ref{#1}}
\newcommand{\algoref}[1]{Algorithm~\ref{#1}}

\newcommand{\secref}[1]{Sect.~\ref{#1}}
\newcommand{\lemmaref}[1]{Lemma~\ref{#1}}
\newcommand{\corollaryref}[1]{Corollary~\ref{#1}}

\newcommand{\approach}{}
\def\approach/{CRDT Paxos}

\newcommand{\myabstract}{
General solutions of state machine replication have to ensure that all
replicas apply the same commands in the same order, even in the
presence of failures. Such strict ordering incurs high synchronization
costs caused by distributed consensus or by the use of a leader.

This paper presents a protocol for \emph{linearizable} state machine
replication of conflict-free replicated data types (CRDTs) that
neither requires consensus nor a leader. By leveraging the properties
of state-based CRDTs---in particular, the monotonic growth of a join
semilattice---synchronization overhead is greatly reduced. As a result,
updates only need a single round trip and modify the state
`in-place' without the need for a log. Furthermore, the message size
overhead for coordination consists of a single counter per message.
For queries, we guarantee finite writes termination. We show in an experimental
evaluation that more than 99\,\% of queries can be handled in one to three round
trips under highly concurrent accesses.

Our protocol achieves high throughput without auxiliary processes
such as command log management or leader election. Thus, it is
well suited for practical scenarios that need linearizable
access to CRDT data on a fine-granular scale.
\vspace{-0em}
}

\begin{document}

\maketitle

\begin{abstract}
\myabstract{}
\end{abstract}

\section{Introduction}
The implementation of a replicated state machine (RSM) is a
well-established approach for designing fault-tolerant services. In its
common form, clients submit \textit{update} commands that modify the
state of the replicated object, or \textit{read} commands returning
(part of) its state back to the client.  To guarantee
linearizable~\cite{herlihy1990linearizability} access to an RSM, all
replicas must apply the same commands in the same order. This is
commonly achieved by using a consensus protocol such as
Paxos~\cite{Lamport_2001,Lamport_1998},
Raft~\cite{Ongaro_Ousterhout_2014}, or variations
thereof~\cite{Moraru_Andersen_Kaminsky_2013,Hunt_Konar_Junqueira_Reed,Lamport_2006}.
However, the use of consensus often incurs significant synchronization
overhead. In particular, most approaches require the use of a central
coordinator (leader) to achieve acceptable performance and require to maintain a
command log, which must be regularly truncated to prevent unbounded
memory consumption. This often makes the correct implementation of
RSMs a challenging task~\cite{Chandra_Griesemer_Redstone_2007}.

A wealth of previous work exists that aims to reduce the cost
associated with fault-tolerant replication. Some approaches reduce
synchronization by leveraging the commutativity of some
submitted commands by solving generalized
consensus~\cite{Lamport_2005}.  Other approaches avoid the cost
associated with consensus by using a weaker consistency model such as
strong eventual consistency (SEC). SEC was formalized by Shapiro et
al.~\cite{Shapiro_Preguica_Baquero_Zawirski_2011} with the
introduction of conflict-free replicated data types (CRDTs). CRDTs are
data structures whose mathematical properties ensure the convergence
of all replicas as long as all updates are propagated to them in arbitrary order.
They do not require protocol-level conflict resolution mechanisms, as conflicting
updates can be resolved computationally. This
allows the conflict-free execution of both queries and updates in
relaxed consistency models like SEC. Data structures that can be
implemented as a CRDT include counters, sets, and certain types of
graphs~\cite{Shapiro_Preguica_Baquero_Zawirski_2011}. Due to their low
synchronization costs, numerous practical systems have employed CRDTs
to this date, such as Redis~\cite{redis_crdt}, Riak~\cite{brown2014riak},
SoundCloud~\cite{soundcloud_roshi}, and Akka~\cite{akka_framework}.

However, their usage is restricted to cases where relaxed consistency suffices, as there is
no guarantee on when replicas converge and inconsistent states can be observed
in the meantime. This prevents their usage to implement, for example,
atomic counters, which is a ubiquitous primitive in distributed computing.

This paper introduces a protocol to implement a special class of
replicated state machines that allows linearizable access on CRDTs
without the need of log management while keeping the message size
overhead at a single counter per message. These RSMs support
\emph{update} operations that modify the state and \emph{query}
operations that return a value but do not modify the state.
Operations that both modify the state and return a value are not
supported.

By leveraging the properties of CRDTs, our protocol can
achieve high throughput even in the absence of a leader. Thereby, the
need for implementing leader election mechanisms is eliminated, which allows
continuous availability as long as a majority of replicas is reachable.
Our protocol does not replicate a log of
commands, which is commonly the case for consensus protocols. Instead,
we replicate the state directly and update it `in-place'. Our protocol
just needs a single counter per replica and it avoids the complexity associated
with command log state and memory management.

Our approach relies on solving \emph{generalized lattice agreement} (GLA). Similar to CRDTs,
values proposed in GLA belong to a join semilattice---a partially ordered set that defines a
join (least upper bound) for all element pairs.
In contrast, for generalized consensus it is not required that such a join always exists.
This difference makes generalized lattice agreement an easier problem to solve.
In fact, previous work has shown that wait-free~\cite{herlihy1991wait} solutions to this problem
exist~\cite{Falerio_Rajamani_Rajan_Ramalingam_Vaswani_2012}, which is proven to
be impossible for consensus~\cite{Fischer_Lynch_Paterson_1985} in an
asynchronous system in the presence of process failures.
However, the protocol described by Faleiro et al.~\cite{Falerio_Rajamani_Rajan_Ramalingam_Vaswani_2012}
requires sending an ever-increasing set of commands in its messages to provide
state machine replication.
In contrast, our approach features message sizes that are bounded by the state of the CRDT
and guarantees finite writes termination~\cite{DBLP:conf/podc/AbrahamCKM04}, which is a weaker termination property than wait-freedom.

The main contributions of this paper are as follows (a brief
announcement of this paper is published in~\cite{crdt-podc-brief}):
\begin{itemize}
    \item We present a protocol that provides linearizable state machine replication
        of state-based CRDTs by solving generalized lattice agreement. The protocol is
        light-weight as it does \emph{not} rely on auxiliary processes for leader election or log management (see \secref{sec:design}).
    \item The protocol processes updates in a single round-trip. Queries support finite
    	writes termination (see \secref{sec:liveness}). We show in our evaluation
    	that more than 99\,\% of queries can be processed in one to three
        round-trips in the presence of a continuous stream of updates (see \secref{sec:evaluation}).
    \item We compare the performance of our protocol with open-source implementations of Paxos
        and Raft, two well-known approaches for linearizable
        RSMs  (see \secref{sec:evaluation}).
\end{itemize}

\section{Preliminaries}
In this section, we discuss the assumed system model and give an introduction to CRDTs.

\subsection{System Model}
We consider a distributed system of $N$ independent and asynchronous
processes $\propSet = \{p_1,p_2,\ldots,p_N\}$, which communicate by message passing.
We consider processes that fail under the crash-stop model and assume unreliable message
transfer, i.e., messages can arrive out of order, can be delayed arbitrarily, or
can be lost. We refer to a process that does not fail as a \textit{correct} process.

We assume over $\propSet$ a fixed quorum system
$\mathit{QS}$~\cite{Vukolic_2012}, i.e., a set of sets of processes with mutual overlap:
\begin{align*}
\forall Q \in \mathit{QS}&: Q \subseteq \propSet\\
\forall Q_1, Q_2 \in \mathit{QS}&: Q_1 \cap Q_2 \neq \emptyset
\end{align*}

Elements in $\mathit{QS}$ are called \textit{quorums}. A necessary condition for progress
is that at least a quorum of processes does not crash and is able
to pairwise exchange messages for a sufficiently long time.

\subsection{State-Based Conflict-Free Replicated Data Types} \label{sec:crdt_intro}
Eventual consistency promises better performance and availability in large scale systems in which the
coordination required for linearizable approaches is not feasible~\cite{viotti2016consistency}.
Updates are applied at some replica and at a later time propagated across the system.
Eventually, all replicas receive all updates, possibly in different orders.
However, concurrent updates may cause conflicts. Resolving
them often requires roll-backs and consensus decisions.

The use of conflict-free replicated data types (CRDTs)~\cite{Shapiro_Preguica_Baquero_Zawirski_2011},
introduced as part of the strong eventual consistency model, eliminates the need for
roll-backs or consensus by leveraging mathematical properties preventing the emergence of conflicts. \emph{Operation-based}
CRDTs require the commutativity of all its update operations, whereas \emph{state-based} CRDTs
rely on monotonicity in a join semilattice~\cite{shapiro2011comprehensive}.
Both types have advantages and disadvantages.
In general, operation-based CRDTs have lower bandwidth needs but
require reliable, i.e., exactly once, and causally ordered delivery of updates~\cite{shapiro2011comprehensive}.
As our system model assumes unreliable communication, we only focus on \emph{state-based
CRDTs} in this paper. However, both types of CRDTs can emulate each other~\cite{shapiro2011comprehensive}.

State-based CRDTs are based on the concept of join semilattices:
\begin{mydef}[Join Semilattice]
    A join semilattice $\mathcal{S}$ is a set $S$ equipped with a
    partial order $x \sqsubseteq y$ and
    a least upper bound (LUB) $x \sqcup y$ for all pairs of elements $x, y \in S$.
\end{mydef}
The LUB of two elements $x,y \in S$ is the smallest element in $S$
that is equal or larger than both $x$ and $y$.
\begin{mydef}[Least Upper Bound]
    $m = x \sqcup y$ is a LUB of $\{x,y\}$ under partial order $\sqsubseteq$ iff:
    \begin{equation*}
        \forall m'\in S,\ x\sqsubseteq m'\land y \sqsubseteq m':
        \quad x \sqsubseteq m\ \land\ y \sqsubseteq m\ \land\ m\sqsubseteq m'
    \end{equation*}
\end{mydef}
From this definition it follows that $\sqcup$ is idempotent ($x \sqcup x = x$),
commutative ($x \sqcup y = y \sqcup x$), and associative
($(x\sqcup y) \sqcup z = x\sqcup(y\sqcup z)$).

The join semilattice represents the set of possible states of a state-based CRDT. Clients
can read its current state via \emph{query} commands and modify it via \emph{update}
commands\footnote{To be consistent with RSM terminology, we use the term 'command'
instead of 'function', which is commonly used in the context of CRDTs.}.
\begin{mydef}[State-Based CRDT]
    A state-based CRDT consists of a triple $(\mathcal{S}, Q, U)$, where $\mathcal{S}$ is a join
    semilattice defining the possible payload states $S$,
    $Q$ is a set of side-effect free query commands, and $U$ is a set of
    monotonically non-decreasing update commands, i.e., $\forall u \in U, s \in S:
    s \sqsubseteq u(s)$. 
\end{mydef}
Two payload states $s_1, s_2 \in S$ are \textit{equivalent} ($s_1
\equiv s_2$) if all queries return the same result for both, i.e.,
$s_1 \sqsubseteq s_2 \land s_2 \sqsubseteq s_1 \implies s_1 \equiv
s_2$.  They are \textit{comparable} if they can be ordered, i.e., $s_1
\sqsubseteq s_2 \lor s_2 \sqsubseteq s_1$.

\begin{algorithm}[t]
    \caption{State-based G-counter replicated on $n$ processes as (non-linearizable) CRDT.}
    \label{fig:gcounter_example}

      \begin{algorithmic}[1]
      \State $S := \mathbb{N}^n$,
      $\sqsubseteq :=$ \textbf{compare},
      $\sqcup := $ \textbf{merge},
      \State $Q := \{\mathrm{\textbf{query}}\}$,
      $U := \{\mathrm{\textbf{update}}\}$
      \funspace
        \Compare
            \State return $\bigwedge_{i=0}^{n-1} x[i] \leq y[i]$
        \EndCompare
        \Merge
            \State $z[i]_{i=0}^{n-1} \gets max(x[i],y[i])$; return $z$
            \vspace*{1ex}
        \EndMerge


        \Payload{$g \in S = [0,\ldots,0]$}\customcomment{G-counter view of replica}
        \EndPayload
        \Query{}{$\mathbb{N}$}\customcomment{get G-counter value of a replica}
        \State return $\sum_{i=0}^{n-1} g[i]$
        \EndQuery
        \Update{}\customcomment{increment G-counter value of a replica}
        \State $i \gets$ my\_replica\_id()
            \State $g[i] \gets g[i] + 1$
        \EndUpdate
        \end{algorithmic}


\end{algorithm}
\medskip
\emph{Example.} One of the most simple state-based CRDTs is a monotonically increasing counter,
called G-counter (grow-only counter). Its state-based definition is shown in
\algoref{fig:gcounter_example}. The payload state of such a counter, replicated
on $n$ processes, consists of an array of length $n$. All replicas, which are assumed to be
distinguishable by an ID, manage their own local copy of the counter's state.
Locally incrementing the counter increments the array element corresponding
to the ID of the respective replica. The \textit{merge} and \textit{compare}
functions implement $\sqcup$ and $\sqsubseteq$, respectively.

In a system that provides SEC, a replica that receives an increment command from
a client increments its counter (its slot) by calling \textit{update}. It periodically
propagates its counter state $g$ to the other replicas. Any replica that receives
such a counter state updates its own counter state using \textit{merge}.
As all replicas only increment their own slot, no updates are lost and eventually all
replicas converge to the same state.

\section{Linearizable and Logless RSM \\of State-Based CRDTs} \label{sec:design}
Next, we discuss how to leverage the properties of state-based CRDTs for linearizable access.

\subsection{Problem Statement} \label{sec:problem_statement}
We consider a state-based CRDT $(\mathcal{S},Q,U)$ replicated on $N$ processes. Each process
starts with an initial state $s_0 \in S$. Clients can perform update and
query operations by respectively sending an update command $u \in U$ or query command
$q \in Q$ to any process. Each process may receive an arbitrary number of commands.
An operation is \emph{invoked} if a client sends a corresponding message to a process,
which may eventually \emph{respond} by sending a message with the operation result back.
An operation $op_1$ \emph{precedes} $op_2$, if a process sends an $op_1$ response before
$op_2$ is invoked. For brevity, we refer to query and update commands as queries and updates.
Furthermore, a command is invoked if the operation it is included in is invoked. Command precedence
is defined analogously.

Updates modify the state of the CRDT without returning a result to the
client (besides a completion acknowledgment). To simplify formal reasoning
we assume updated to be unique, e.g., by attaching IDs.
The \emph{causal history}~\cite{Shapiro_Preguica_Baquero_Zawirski_2011} $C(s)$ of
state $s$ is the set of updates applied on $s_0$ to reach state
$s$. More formally, $C(s_0) = \emptyset$, $u \in C(u(s))$ and $C(s_1)
\cup C(s_2) = C(s_1 \sqcup s_2)$. Shapiro et
al.~\cite{Shapiro_Preguica_Baquero_Zawirski_2011} have shown that
$C(s_1) = C(s_2) \Rightarrow s_1 \equiv s_2$ due to the properties of
updates and LUBs.  A state $s$ \emph{includes} update $u$ if $u \in
C(s)$.  Note that causal histories are an aid for formal reasoning, which
do not have to be explicitly stored by an implementation.

In contrast to updates, queries do not modify the state of the CRDT but return a value
as result. To process a query $q$ that was sent to a
process $p$, $p$ must first \textit{learn} a
state $s \in S$ by exchanging messages with the other processes. The query is
then applied on $s$ and the result is returned to the client. We say that $s$ is the state learned by query $q$ at process $p$.

All learned states must satisfy the following conditions:

\begin{description}
    \item [Validity] The causal history of any learned state is a subset of all previously invoked updates.
    \item [Stability] For any two states $s_1, s_2$ learned by
        queries $q_1, q_2$, where $q_1$ precedes $q_2$: $s_1 \sqsubseteq s_2$.
    \item [Consistency] Any two learned states are comparable.
\end{description}

These conditions are derived from generalized lattice agreement (GLA)~\cite{Falerio_Rajamani_Rajan_Ramalingam_Vaswani_2012}.
Informally, they capture the notion that queries observe the effect of
a monotonically increasing set of invoked updates.

The conditions stated above define the behavior of queries. We now
define the behavior of updates.

\begin{description}
    \item [Update Stability]  If update $u_1$ precedes update $u_2$,
        then every learned state that includes $u_2$ also includes $u_1$.
    \item [Update Visibility] If update $u$ precedes query $q$, then the
        state learned by $q$ includes $u$.
\end{description}

We show in \secref{sec:linearizability} that these condition suffice to
provide linearizability.

\subsection{The Protocol} \label{sec:protocol}
The success path of the protocol is depicted in \algoref{fig:code_crdt_paxos}.
We consider two roles that processes can assume: \textit{proposer} and
\textit{acceptor}. Roughly speaking, proposers process incoming requests from
clients and acceptors act as the replicated storage of the CRDT. We assume that
all processes implement both the acceptor and proposer role.

\begin{algorithm*}[t]
        \caption{Linearizable state machine replication of state-based CRDTs.}
        \label{fig:code_crdt_paxos}
        
    \setlength{\columnseprule}{0.4pt}
    \begin{multicols}{2}[\vspace*{-3ex}]
      \noindent
      \begin{algorithmic}[1]


            \Statex Proposer:
            \Statex\hdashrule[0.4ex]{.34\linewidth}{0.5pt}{3pt}%
            \emph{Update Commands\hfill}
            \hdashrule[0.4ex]{.34\linewidth}{0.5pt}{3pt}

            \Receive{\mathit{UPDATE}, \mathit{cmd}_u}{client $c$}
                \State store $c$
                \State $s \gets$ apply\_update($\mathit{cmd}_u$) \customcomment{called on local acceptor}
                \State send \msg{\mathit{MERGE}, s} to remote acceptors
            \EndReceive

            \funspace

            \Receive{\mathit{MERGED}}{a quorum}
                \State send \msg{\mathit{UPDATE\_DONE}} to $c$
                \vspace*{-1.5ex}
            \EndReceive

            \funspace
            \funspace
            \Statex\hdashrule[0.4ex]{.35\linewidth}{0.5pt}{3pt}%
            \emph{Query Commands\hfill}
            \hdashrule[0.4ex]{.35\linewidth}{0.5pt}{3pt}

            \Receive{\mathit{QUERY, \mathit{cmd}_q}}{client $c$}
                \State store c, $\mathit{cmd}_q$
                \State $r \gets (\bot, \mathit{new\_id()})$ \label{line:round} \customcomment{incremental prepare}
                \State send \msg{\mathit{PREPARE, r, s_0}} to acceptors
            \EndReceive

            \funspace

            \Receive{\mathit{ACK, \breve{R}, \breve{S}}\label{line:prepare_done}} {a quorum}
                \State $s' \gets \sqcup \breve{S}$\label{line:lub_in_ack} \customcomment{merge received states}
                \If{$\forall s_i \in \breve{S}: s_i \equiv s'$} \label{line:cons_learned_start}
                    \State $\triangleright$ \emph{$s'$ learned by consistent states}\label{line:cons_learned_state}
                    \State send \msg{\mathit{QUERY\_DONE}, \mathit{cmd}_q(s')} to c \label{line:cons_learned_end}
                \ElsIf {$\forall r_i, r_j\in \breve{R}: r_i = r_j$} \label{line:propose_start}
                    \State $\triangleright$ \emph{consistent rounds}
                    \State send \msg{\mathit{VOTE, \text{any}\ r \in \breve{R}, s'}} to acceptors
                        \label{line:send_vote}
                        \label{line:propose_end}
                \Else
                    \State $\triangleright$ \emph{inconsistent rounds, retry with larger r}
                    \State $r' \gets max(\breve{R})$ \label{line:retry_start}
                    \State $r \gets (r'_{\mathit{nr}} + 1, new\_id())$
                    \State send \msg{\mathit{PREPARE}, r, s'} \label{line:retry_end}
                \EndIf
            \EndReceive

        \funspace

        \Receive{\mathit{VOTED, \{s,\ldots,s\}}}{a quorum}
            \State $\triangleright$ \emph{$s$ learned by vote}
            \label{line:cons_voted}
            \State send \msg{\mathit{QUERY\_DONE}, \mathit{cmd}_q(s)} to c
        \EndReceive

            \columnbreak

            \Statex Acceptor:
            \Init
            \State $r \gets (0,\bot)$
                \State $s \gets s_0$\label{line:acc_init_s}
                \vspace*{-1.5ex}
            \EndInit

            \funspace

            \Statex\hdashrule[0.4ex]{.34\linewidth}{0.5pt}{3pt}%
            \emph{Update Commands\hfill}
            \hdashrule[0.4ex]{.34\linewidth}{0.5pt}{3pt}

            \funspace

            \Fun{apply\_update}{$\mathit{cmd}_u$}
                \State $s \gets \mathit{cmd}_u(s)$
                \State $r_{\mathit{id}} \gets \bot$ \customcomment{invalidate round in progress (see line~\ref{line:vote_cond})}
                \State \Return $s$
            \EndFun

            \funspace

            \Receive{\mathit{MERGE}, s'}{proposer $p$}
                \State $s \gets s \sqcup s'$
                \State $r_{\mathit{id}} \gets \bot$ \customcomment{invalidate round in progress (see line~\ref{line:vote_cond})}
                \State send \msg{\mathit{MERGED}} to $p$
                \vspace*{-1.5ex}
            \EndReceive

            \funspace

            \Statex\hdashrule[0.4ex]{.35\linewidth}{0.5pt}{3pt}%
            \emph{Query Commands\hfill}
            \hdashrule[0.4ex]{.35\linewidth}{0.5pt}{3pt}
            
            \funspace

            \Receive{\mathit{PREPARE}, r', s'}{proposer $p$}
                \State $s \gets s \sqcup s'$
                    \label{line:acc_prepare_update_s}
                \If {$r'_{\mathit{nr}} = \bot$}
                \State $r' \gets (r_{\mathit{nr}}+1, r_{\mathit{id}}')$ \label{line:incremental_prep} \customcomment{set $r'_{nr}$ based on local $r_{nr}$}
                \EndIf
                \If {$r_{\mathit{nr}}' > r_{\mathit{nr}}$} \label{line:fixed_prep_start}
                    \State $r \gets r'$\label{line:acc_update_round}
                    \State send \msg{\mathit{ACK}, r, s}\label{line:acc_send_ack}
                \EndIf \label{line:prep_end}
            \EndReceive

            \funspace

            \Receive{\mathit{VOTE}, r', s'}{proposer $p$}
                \State $s \gets s \sqcup s'$ 
                    \label{line:acc_vote_update_s}
                \If{$r' = r$}\label{line:vote_cond}
                \customcomment{same round as latest PREPARE?}

                    \State $r \gets r'$
                    \State send \msg{\mathit{VOTED, s'}}\label{line:send_voted}
                \EndIf
            \EndReceive
        \end{algorithmic}

    \end{multicols}
    \vspace*{-2.5ex}
    \setlength{\columnseprule}{0pt}

\end{algorithm*}

\begin{description}[listparindent=\parindent, topsep=0.5em, leftmargin=0cm, itemsep=0.5em]
\item[Conventions.]

To keep the presented code brief, we follow several conventions.
First, we assume \textit{messages} to be tuples with a tag and an
arbitrary number of elements. They are denoted as
\msg{\mathit{TAG}, e_0,\ldots, e_n}. Processes wait until they have received
enough messages with a specific tag before executing its corresponding action.
If an action requires messages from a set of processes, we aggregate the received
messages element-wise into multisets. For example, two messages \msg{\mathit{TAG}, a_0, b_0},
\msg{\mathit{TAG}, a_1, b_1} would be aggregated into the message
\msg{\mathit{TAG}, \breve{A} = \{a_0, a_1\}, \breve{B} = \{b_0, b_1\}}.
At any time, each process executes at most one action.

The second concept we use are \textit{rounds}. A common way to
generate unique round numbers is that each process appends its process
ID to a local counter, which is incremented for each new round.  Thus,
rounds are pairs of a round number and a round ID. Round $r$ is
denoted as $r=(\mathit{number}, \mathit{ID})$, with $r_{\mathit{nr}}$
and $r_{\mathit{id}}$ providing access to its number and ID,
respectively.  Round numbers are used to order concurrent requests,
and round IDs guarantee that the round of each request is unique. The special
value $\bot$ denotes empty fields, which is smaller than any other round number or
round ID. Rounds are partially ordered by comparing their round numbers. Round
IDs are only relevant for equality checks.

We furthermore assume that proposers implement a mechanism to keep track of ongoing
requests and can differentiate to which request an incoming message belongs to. In practice,
this can be done by generating a unique ID per request which is included in all messages.

\item[Internal State.]
Each acceptor holds as its internal state the current payload state
$s$ of the CRDT and the highest round $r$ it has observed so far. In the beginning,
each acceptor's state is initialized with some initial payload state $s_0$ and some round
with round number $0$ and an ID that is smaller than any ID generated by proposers.

Proposers only have to temporarily store data of ongoing requests and unprocessed
messages (in order to wait for replies from a quorum). No further state is required.

\item[Update Operations.]
Update operations are processed in a single round
trip. They do not require any synchronization. If a proposer receives an
update command $\mathit{cmd}_u \in U$, it applies the
update locally and sends the resulting new payload state to all other acceptors
in a $\mathit{MERGE}$ message. Upon receiving the message, each acceptor updates
its own payload state by LUB computation and sends an acknowledgment message
back to the proposer. After receiving replies from a quorum, the update is complete
and the client is notified by the proposer.

\item[Query Operations.]
Query operations require synchronization as a quorum
must agree upon some payload state in order to satisfy
Validity, Stability, and Consistency
(\secref{sec:problem_statement}). This is achieved with
a modified variant of the Paxos algorithm~\cite{Lamport_2001}.

\medskip
Proposer $p$ begins the query protocol with the reception of
a query command $\mathit{cmd}_q \in Q$. Before executing the command, it must first learn
the current payload state in two phases. First,
$p$ announces its intent to learn a state with $\mathit{PREPARE}$ messages and
then proposes to learn a state, which acceptors have to agree on.

In the first phase, $p$ first chooses a round (line~\ref{line:round}), which
is later used for the proposal in the second phase.
The round number can be chosen by $p$ in two ways.
    First, $p$ can decide on a fixed integer as a round number. We refer to this as
a \textit{fixed prepare}. The chosen number should be larger than all
round numbers previously chosen by any proposer, as otherwise $p$ cannot succeed
in this phase. However, $p$ has only knowledge of its own proposals, which can
make it difficult to decide on an acceptable number. Therefore, proposers may choose to opt for
an \textit{incremental prepare} by leaving the round number undefined
(denoted as $\bot$).

In addition to a round, $p$ includes its own payload state in its $\mathit{PREPARE}$
message. This state can be either $s_0$, or some recently observed state $s$.
Including such a state is not required for safety, but it can speed-up the convergence
of acceptors' payload states.

Each acceptor updates its
rounds and payload state according to the $\mathit{PREPARE}$
message it receives (incremental or fixed).
Note that acceptors do not accept a fixed prepare if it includes a round with
a round number smaller than the highest round number already
seen by this acceptor (lines~\ref{line:fixed_prep_start}--\ref{line:prep_end}).
In practice, the acceptors reply with $\mathit{NACK}$ messages (not shown for brevity)
so that the proposer can retry its request. An incremental prepare is always
accepted and the local round number of the acceptor is increased
(line~\ref{line:incremental_prep}).

The prepare is successful if a quorum has replied with $\mathit{ACK}$ messages
(line~\ref{line:prepare_done}).
Depending on the replies, $p$ can either (a) immediately learn
a state, (b) propose a state to learn, or (c) retry the prepare phase.

(a) If all acceptors of the quorum replied with the same payload
state, then this state can be considered to be learned by $p$. Thus,
the second phase can be skipped, $p$ can apply $\mathit{cmd}_q$ on the learned
state, and send the result to the client.  We refer to such state as
\textit{learned by consistent quorum}
(lines~\ref{line:cons_learned_start}--\ref{line:cons_learned_end}).
The second phase can be skipped here as $p$ is already
certain of a payload state that is established in a quorum.

(b) If a quorum of acceptors replied with the same
round, the first phase was successful.
In the second phase, the proposer can propose a payload state to learn, which is the
LUB of all received acceptor payloads. This state is sent with the round used in the first
phase in $\mathit{VOTE}$ messages to all acceptors
(lines~\ref{line:propose_start}--\ref{line:propose_end}).

(c) If neither payload states nor rounds are consistent, the first
phase has failed. In this case, the proposer used an incremental prepare. It
can then retry with a fixed prepare by choosing a round number that is larger
than all seen round numbers
(lines~\ref{line:retry_start}--\ref{line:retry_end}).

Each acceptor that received a \msg{\mathit{VOTE}, s', r'} message has
to decide whether the
proposal is valid. This is the case when the acceptor has received $p$'s
$\mathit{PREPARE}$ message and its state was not modified by a concurrent update or query
in the meantime (line~\ref{line:vote_cond}).
If the proposal is valid, then the acceptor replies with a $\mathit{VOTED}$ message.
Otherwise, it denies the proposal by optionally sending a $\mathit{NACK}$ so that
$p$ can retry (not shown). If $p$ receives a quorum of $\mathit{VOTED}$ messages, then its proposed
state is learned. We refer to this as a state \textit{learned by
  vote}. Then, $p$ can apply the received query and send the result to the client.

\item[Retrying Requests.]
Acceptors may deny concurrently submitted queries by sending $\mathit{NACK}$
to the respective proposer. It is helpful to include the current payload state
of the denying acceptor in this message to speed-up the convergence with the remaining acceptors in
the system.

Any proposer that received a $\mathit{NACK}$ before receiving a quorum of $\mathit{ACK}$ or $\mathit{VOTED}$
messages must retry its request. It can compute the LUB of all received payloads as the
state to include in its next $\mathit{PREPARE}$ messages. By always retrying with
an incremental prepare, eventual liveness (see \secref{sec:liveness}) can be
guaranteed. However, retrying with a fixed prepare also does not violate any safety condition of \secref{sec:problem_statement}.

\end{description}

\subsection{Relation to Paxos and ABD}
The query protocol is closely related to the classical single-decree Paxos
algorithm~\cite{Lamport_1998}.
Single-decree Paxos can be used to agree on a single value or
command sent to the RSM. This makes it necessary to use multiple chained Paxos instances
to learn a sequence of commands. Paxos solves consensus for arbitrary values. Therefore, it can
not assume that properties such as commutativity and idempotence generally exist. In contrast,
CRDT state merges always exhibit these properties, which allows us to modify Paxos to exploit them.
First, a single instance of our protocol can be re-used to repeatedly merge states received from
proposers in arbitrary order, even if the attached round number is outdated.
This speeds-up the convergence of acceptors states under concurrent access.
Second, our approach needs only a single round number, whereas Paxos requires an
additional round number in the state of acceptors to identify the newest proposed value in the case of concurrent
proposals. As we can simply merge all observed values, this round number is not needed.
Third, proposers in our approach can terminate early if they observe consistent states from a
quorum of acceptors. This optimization makes our approach viable in
leader-less deployments, as shown in \secref{sec:evaluation}.

Due to the modifications made to Paxos, our approach somewhat resembles the
multi-writer generalization of the ABD algorithm~\cite{DBLP:journals/jacm/AttiyaBD95}.
ABD provides a wait-free fault-tolerant atomic register. As such, newer values
submitted by clients overwrite the old register state in ABD. These semantics alone
do not suffice for state machine replication, which requires
sequential agreement on commands. For our state-based CRDTs, accepted updates
are merged into the previous
state, which ensures that clients observe monotonically increasing CRDT states.

\subsection{Proof of Safety}
In the following, we prove that our protocol satisfies the conditions outlined in \secref{sec:problem_statement}.
The query protocol begins by either incremental or fixed prepare. The following invariants hold for both of them,
as can be directly inferred from \algoref{fig:code_crdt_paxos}:
\begin{description}[style=multiline, leftmargin=4ex]
    \item[I1] If a proposer learns some state, then it has received
        $\mathit{ACK}$'s from a quorum (line~\ref{line:cons_learned_state}, line~\ref{line:cons_voted} via line~\ref{line:send_vote}).
    \item[I2] Any learned state is the LUB of all payload states received in
        $\mathit{ACK}$ messages from a quorum (line~\ref{line:lub_in_ack}).
    \item[I3] If a proposer sends a $\mathit{VOTE}$ message, then it has received
        the same round in $\mathit{ACK}$ messages from a quorum (lines~\ref{line:propose_start}--\ref{line:send_vote}).
    \item[I4] If a proposer has received an $\mathit{ACK}$ message
      from an acceptor (line~\ref{line:prepare_done}),
        then this acceptor has increased its round number due to the proposer's
        $\mathit{PREPARE}$ message (line~\ref{line:acc_update_round}).
\end{description}

\begin{theorem}[Validity] \label{th:validity}
    The causal history of any learned state is a subset of all previously invoked updates.
\end{theorem}
\begin{proof}
All acceptors start with payload $s_0$ (line~\ref{line:acc_init_s}). Payload modifications
only happen by either application of a received update command or by LUB computation.
Every update is applied at most once and only if it was previously received by a proposer.
Furthermore, computing the LUB of two payload states computes the union of their respective
causal histories. Thus, the causal history of every learned state must be a subset
of previously invoked updates.
\end{proof}

\begin{lemma}\label{lemma.monoton}
    The payload state of every acceptor increases monotonically.
\end{lemma}
\begin{proof}
Both LUB computation and the direct application of update commands are monotonically
increasing.
\end{proof}

\begin{corollary} \label{proof:corollary}
    If messages \msg{\mathit{ACK}, r, s} and \msg{\mathit{ACK}, r', s'}
    are send by the same acceptor in this order, then $s \sqsubseteq s'$.
\end{corollary}

\begin{lemma} \label{proof:lemma_quorum_larger}
    If state $s$ is learned by any proposer, then there exists a quorum $Q$ with
    $s \sqsubseteq a.s, \forall a \in Q$, where $a.s$ designates the
    local state variable $s$ of an acceptor process $a$.
\end{lemma}
\begin{proof}
State $s$ can be learned (i) by consistent quorum from messages of a
quorum $Q_{cons}$ (line~\ref{line:cons_learned_state}) or
(ii) by vote from messages of a quorum $Q_{\mathit{vote}}$ (line~\ref{line:cons_voted}).
\begin{description}[style=multiline, leftmargin=4.3ex]
    \item[(i)] Trivial, as all acceptors in $Q_{\mathit{cons}}$ have included
        a state $s' \equiv s$ in their $\mathit{ACK}$ message
        (lines~\ref{line:acc_prepare_update_s} and~\ref{line:acc_send_ack}).
    \item[(ii)] $p$ sent $s$ in $\mathit{VOTE}$ messages. At least all acceptors in
        $Q_{\mathit{vote}}$ must have received the message and have merged their payload state
        with $s$ by LUB computation (line~\ref{line:acc_vote_update_s}) before replying with $\mathit{VOTED}$ (line~\ref{line:send_voted}).
\end{description}
\end{proof}

\begin{theorem}[Stability]  \label{th:stability}
  For any two states $s_1, s_2$ learned by queries $q_1, q_2$, where
  $q_1$ precedes $q_2$: $s_1 \sqsubseteq s_2$.
\end{theorem}
\begin{proof}
From \lemmaref{proof:lemma_quorum_larger} it follows that once a proposer $p$
has received the $\mathit{QUERY}$ message of $q_2$, there exists a quorum $Q$ such
that $s_1 \sqsubseteq a.s, \forall a \in Q$. To learn a state, $p$ eventually
receives $\mathit{ACK}$ messages from quorum $Q'$. As $Q \cap Q' \neq \emptyset$,
there exists some $a'\in Q'$ with $s_1 \sqsubseteq a'\!\!.s$.  The state learned
by $p$ is the LUB of all received states included in the $\mathit{ACK}$ messages.
Thus, $s_1 \sqsubseteq a'\!\!.s \sqsubseteq s_2$.
\end{proof}

\begin{lemma} \label{proof:lemma_cons_a}
    Two learned states $s_1$ and $s_2$ are comparable if at least one state is
    learned by consistent quorum.
\end{lemma}
\begin{proof} (By contradiction)
Let $s_1$ and $s_2$ be learned due to queries handled at proposer
$p_1$ and $p_2$, respectively. $p_1$ and $p_2$ have received
$\mathit{ACK}$s from quorums $Q_1$ and $Q_2$, respectively. Assume
$s_1$ is learned by consistent quorum and $s_1$ is not comparable to $s_2$.
In this case, the following conditions must hold:
    \begin{description}[style=multiline, leftmargin=4ex]
        \item[C1] $\forall a \in Q_1 \cap Q_2$: $a$ must send an $\mathit{ACK}$ to $p_2$ with
            state $s: (s \sqsubseteq s_1) \land \neg (s \equiv s_1)$, otherwise $s_1 \sqsubseteq s_2$. This implies that
            $a$ receives $p_2$'s $\mathit{PREPARE}$ message before $p_1$'s.
        \item[C2] $\forall a \in Q_1$: $a$ must receive a
            $\mathit{PREPARE}$ message from $p_1$ before receiving
            $\mathit{VOTE}$ from $p_2$ (otherwise $s_2 \sqsubseteq s_1$).
    \end{description}

    $s_2$ cannot be learned by consistent quorum, as this would imply
    $s_2 \equiv s \sqsubseteq s_1$ (C1 and \corollaryref{proof:corollary}). Thus, to learn $s_2$,
    $p_2$ must receive $\mathit{VOTED}$ messages from a quorum with at least one acceptor $a$ in $Q_1$.
    For that, $p_2$ sends a \msg{\mathit{VOTE}, r, s_2} message to $a$.
    It follows from C1 and C2 that $a$ has received $p_1$'s
    $\mathit{PREPARE}$ message in between $p_2$'s $\mathit{PREPARE}$ and $\mathit{VOTE}$ message.
    Due to invariant I4, $a$ has modified its round and $r \neq a.r$. Therefore, $a$
    does not reply with a $\mathit{VOTED}$ message and $s_2$ cannot be learned.
\end{proof}

\begin{lemma} \label{proof:lemma_cons_b}
    Two learned states $s_1$ and $s_2$ are comparable if both are learned by
    vote.
\end{lemma}
\begin{proof}
Let $s_1$ and $s_2$ be learned due to query requests handled at proposer $p_1$ and
$p_2$, respectively. $p_1$ has received $\mathit{ACK}$s from quorum $Q_1$ and $p_2$ from quorum
$Q_2$. As $Q_1 \cap Q_2 \neq \emptyset$, there is at least one
acceptor $a$ that has
sent $\mathit{ACK}$s to both $s_1$ and $s_2$. Assume $a$ sends an $\mathit{ACK}$ to $p_1$ first. Therefore,
    $p_1$ sends \msg{\mathit{VOTE}, r_1, s_1} and $p_2$ sends \msg{\mathit{VOTE}, r_2, s_2} with
    $r_1 < r_2$. Let $Q_v$ be the quorum of acceptors that replied to
    $p_1$ with $\mathit{VOTED}$ messages. All acceptors $a \in Q_v \cap Q_2$
    must receive $p_1$'s $\mathit{VOTE}$ before $p_2$'s $\mathit{PREPARE}$
    message, as otherwise either $p_2$ receives inconsistent rounds
    or $a$ does not reply to $p_1$. Therefore, $a$ includes state $s$ with $s_1 \sqsubseteq s$
    in its $\mathit{ACK}$ message to
    $p_2$. As $p_2$ computes the LUB of all states received
    in $\mathit{ACK}$ messages, $s_1 \sqsubseteq s \sqsubseteq s_2$.
\end{proof}
\begin{theorem}[Consistency] \label{th:consistency}
    Any two learned states are comparable.
\end{theorem}
\begin{proof}
    Follows from \lemmaref{proof:lemma_cons_a} and \lemmaref{proof:lemma_cons_b}.
\end{proof}

\begin{theorem}[Update Stability] \label{th:ustability}
    If update $u_1$ precedes update $u_2$, then every learned state that includes
    $u_2$ also includes $u_1$.
\end{theorem}
\begin{proof}(By contradiction)
    As $u_1$ precedes $u_2$, there is a quorum $Q_u$ that has
    received $\mathit{MERGE}$ messages with a payload including $u_1$
    before any acceptor includes $u_2$. Thus, there cannot be a quorum at
    any time that includes $u_2$ but not $u_1$.

    Assume a proposer $p$ learns state $s$ that includes $u_2$
    but not $u_1$. So, there must be a quorum
    $Q_{ack}$ that
    has replied to $p$ an $\mathit{ACK}$ message before receiving the $\mathit{MERGE}$
    message and at least one acceptor replied with a payload including $u_2$.
    It follows that $s$ is not learned by consistent quorum. It also follows that
    all acceptors in $Q_u$ received the $\mathit{MERGE}$ before $p$ received all
    replies from $Q_{ack}$. To propose a state in $\mathit{VOTE}$
    messages, $p$ must have received the same round $r$ from all acceptors in
    $Q_{ack}$. However, $\nexists Q: a.r = r, \forall a \in Q$, as $\forall a \in Q_u$
    updated their round. Therefore, $p$'s proposal cannot succeed and $s$ is not learned by vote.
\end{proof}

\begin{theorem}[Update Visibility] \label{th:uvisibility}
    If update $u$ precedes query $q$, then the state learned by $q$ includes $u$.
\end{theorem}
\begin{proof}
    Since the proposer processing update $u$ has generated a response event,
    there exists a quorum of
    acceptors including $u$. Thus, any proposer that processes a subsequent
    query receives at least one $\mathit{ACK}$ message that includes $u$.
\end{proof}

\subsection{Proof of Linearizability} \label{sec:linearizability}
\newcommand{\uSet}{\mathcal{U}}
\newcommand{\Hist}{H}
\newcommand{\HExt}{\bar{H}}
\newcommand{\HSeq}{S}
\newcommand{\precH}{\prec_{\Hist}}
\newcommand{\precHExt}{\prec_{\HExt}}
\newcommand{\precHSeq}{\prec_{\HSeq}}
\newcommand{\precInc}{\prec_{Inv}}

In this section, we show that any protocol that satisfied Theorems~1--5 provides linearizable
access to CRDTs. Therefore, it may be of general interest, independent
of our proposed protocol.
In the sequential specification of a CRDT, the state $q.s$ learned by
query $q$ satisfies the following condition. Let $\uSet_q$ be the set of all
updates that precede $q$.
\begin{equation} \label{eq:seq_spec}
  \exists s' \equiv q.s: C(s') = \uSet_q
\end{equation}
Informally, this captures the notion that the effect of all preceding
updates must be observed by the query. The effect of future updates
must not be visible.

Furthermore, we rely on the following Lemma.
\begin{lemma} \label{lm:lin}
  $s_1 \sqsubseteq s_2 \Rightarrow \exists s \equiv s_2: C(s_1) \subseteq C(s)$
\end{lemma}
\begin{proof}
  Because of $s_1 \sqcup s_2 = s \equiv s_2$ and $C(s_1 \sqcup s_2) = C(s) = C(s_1) \cup C(s_2)$,
  it follows that $C(s_1) \subseteq C(s)$.
\end{proof}

Let $\precH$ be an irreflexive partial order on operations induced by
history $\Hist$~\cite{herlihy1990linearizability}. Then, $op_1 \precH
op_2$ is satisfied if $op_1$ precedes $op_2$.
In the following, we denote query and update operations by $r$ and $w$, respectively.
For example, $\forall r \in_r X$, denotes 'for all query operations in X'. Operations that can be of either
type are denoted by $o$, where $o.f$ is the command of operation $o$.
We denote the
state learned by query operation $r$ as $r.s$.
\begin{proof}
  Fix an execution $E$ of an algorithm that satisfies Theorems~1--5 with history $\Hist$. Let
  $\HExt$ be an extension of $\Hist$ by adding response events to pending invocations. Let
  $\precInc$ be a total order over $\HExt$, where $o_1 \precInc  o_2$ if the invocation
  of $o_1$ precedes the invocation of $o_2$.
  We construct a sequential history $\HSeq$ from $\HExt$. For every pair $(o_1, o_2)$, $o_1 \neq o_2$
  in $\HExt$:
\begin{itemize}
  \item[S1] For $(w, r)$ or $(r, w)$:\\
    if $r \not\precHExt w\ \land\ \exists s' \equiv r.s: w.f \in C(s') \Rightarrow w \precHSeq r$\\
    else $r \precHSeq w$
  \item[S2] For $(r_1, r_2)$:\\
    $r_1.s \sqsubset r_2.s\ \lor\ (r_1.s \equiv r_2.s\ \land\  r_1
    \precInc r_2) \Rightarrow r_1 \precHSeq r_2$
  \item[S3] For $(w_1, w_2)$:\\
      Let $r^* \notin \HSeq$ be an auxiliary query operation after $w_1$ and
      $w_2$ with $\forall r \in \HSeq: r \precHSeq r^*$; The queries
      immediately following $w_1$ and $w_2$ are then:\\
      $r_{w_i} = \min_{\precHSeq}(\{r \in_r \HSeq: w_i \precHSeq r\} \cup \{r^*\}),\  i \in \{1,2\}$;\\
      $r_{w_1} \precHSeq r_{w_2}\ \lor\ (r_{w_1} = r_{w_2} \ \land\ w_1 \precInc w_2) \Rightarrow w_1 \precHSeq w_2$
\end{itemize}
The set of query operations in $\HSeq$ is totally ordered by $\precHSeq$ due to Theorem~\ref{th:consistency}.
Thus, $\min_{\precHSeq}$
in case S3 is well defined and always exists. It is easy to
see that $\precHSeq$ is antisymmetric, i.e., $(o_1 \precHSeq o_2) \oplus
(o_2 \precHSeq o_1)$, for $o_1 \neq o_2$, where $\oplus$ denotes
the exclusive or operator.

\begin{lemma}
  $\precHExt \subseteq \precHSeq$
\end{lemma}
Theorems~\ref{th:validity}, \ref{th:stability}, \ref{th:ustability}, and \ref{th:uvisibility}
define the behavior of any pair of non-overlapping operations.
Let $o_1, o_2 \in \HExt$. If $o_1$ and $o_2$ are query operations,
$o_1 \precHExt o_2 \Rightarrow o_1 \precHSeq o_2$ follows trivially by
case S2 and Theorem~\ref{th:stability}. If either $o_1$ or $o_2$ is an update, the same follows directly
by case S1 and Theorem~\ref{th:uvisibility} or \ref{th:validity}, respectively.
Let both $o_1$ and $o_2$ be update operations. Theorem~\ref{th:ustability} states:
\begin{equation} \label{eq:subset}
o_1 \precHExt o_2 \Rightarrow \forall r \in_r \HExt, o_2.f \in C(r): o_1.f \in C(r)
\end{equation}
Let $r_{w_1}$ and $r_{w_2}$ be the query operations following $o_1$ and $o_2$ respectively, as defined in case S3. It follows from
equation~\ref{eq:subset} that $o_1.f \in C(r_{w_2})$. Thus, $r_{w_2} \not\precHSeq r_{w_1}$. Therefore,
$r_{w_1} \precHSeq r_{w_2}$ or $r_{w_1} = r_{w_2}$. In both cases, $o_1 \precHSeq o_2$.

\begin{lemma}
  $\HSeq$ is a sequential history.
\end{lemma}
The relation $\precHSeq$ is antisymmetric and relates all pairs of elements in some way.
Thus, $\precHSeq$ is a total order if $\precHSeq$ is also transitive, which implies
that $\HSeq$ is a sequential history. Let $r_1, r_2, r_3$ and
$w_1, w_2, w_3$ be any three query or
update operations, respectively.

(A) $r_1 \precHSeq r_2 \precHSeq r_3 \Rightarrow r_1 \precHSeq r_3$,
follows trivially from case S2.
In addition, case S2 implies $r_1 \precHSeq r_2 \Rightarrow r_1.s \sqsubseteq r_2.s$. Thus, by applying
Lemma~\ref{lm:lin} and case S1, (B) $w_1 \precHSeq r_1 \precHSeq r_2 \Rightarrow w_1 \precHSeq r_2$.
Let $r_{w_1}$ and $r_{w_2}$ be the query operations following $w_1$ and $w_2$ respectively, as defined in case S3.
From (A) and case S3 follows $w_1 \precHSeq w_2 \Rightarrow r_{w_2} \not\precHSeq r_{w_1}$,
thereby (C) $w_1 \precHSeq w_2 \precHSeq r_1 \Rightarrow w_1 \precHSeq r_1$
and (D) $w_1 \precHSeq w_2 \precHSeq w_3 \Rightarrow w_1 \precHSeq w_3$.

We derive from (B) that (E) $r_1 \precHSeq r_2 \precHSeq w_1 \Rightarrow r_1 \precHSeq w_1$ holds
using the following argument: Assume $r_1 \not\precHSeq w_1$. By antisymmetry of $\precHSeq$, $w_1 \precHSeq r_1$.
We arrive at an contradiction because $r_1 \precHSeq r_2$,
which implies $w_1 \precHSeq r_2$ due to (B).

As this argument does not rely on the operation type, it can be applied to derive
the remaining cases from (C) and (E). Thereby, $\precHSeq$ is transitive.

\begin{lemma}
  $\HSeq$ is legal in respect to the sequential specification of CRDTs.
\end{lemma}
We show that all query operations in $\HSeq$ satisfy equation~\ref{eq:seq_spec} by construction rule S1.
Let $r$ be a fixed query operation in $\HSeq$ and
$\uSet_r = \{w.f: w \in_w \HSeq \land w \precHSeq r\}$, i.e., the set
of all updates preceding $r$.
By case S1:
\begin{equation} \label{eq:u_vis}
  \forall u \in \uSet_r, \exists s' \equiv r.s: u \in C(s')
\end{equation}
Let $\uSet_r = \{u_1,\dots, u_n\}$. We proceed by induction to show that an $s'$
exists, so that $C(s') = \uSet_r$. Let $s_i$ be a state equivalent to $r.s$ that includes all
$u_j$ with $j \leq i$. We show that $s_i$ can be constructed using $s_{i-1}$, starting
$s_0 = r.s$:
\begin{enumerate}
  \item If $u_i \in C(s_{i-1})$, then $s_i = s_{i-1}$.
  \item Otherwise, $s_i = u_i(s_{i-1})$. As updates are non-decreasing, $s_{i-1} \sqsubseteq s_i$.
    Let $s'$ be any state
    that satisfies equation~\ref{eq:u_vis} in respect to $u_i$ and $s_m = s' \sqcup s_{i-1}$. Because of
    $C(s_i) = C(s_{i-1}) \cup \{u_i\} \subseteq C(s_m)$ it follows that $s_m \not\sqsubset s_i$.
    As $s_m \equiv r.s \equiv s_{i-1}$, we have $s_{i-1} \not\sqsubset s_i$. Thus,
    $s_i \equiv s_{i-1}$.
\end{enumerate}
By construction of $s_n$, we know that $C(s_n) = \uSet_r \cup C(r.s)$. By Theorem~\ref{th:validity}
and case S1 it follows $C(r.s) \subseteq \uSet_r$.
Therefore, $C(s_n) = \uSet_r$.

As $\HSeq$ satisfies Lemma 6, 7 and 8, $\HExt$ is linearizable. Since $\HExt$ is
the extention of $\Hist$, $\Hist$ is also linearizable.
\end{proof}

\subsection{Liveness} \label{sec:liveness}
Faleiro et al.~\cite{Falerio_Rajamani_Rajan_Ramalingam_Vaswani_2012} show that wait-free
protocols for solving GLA exist. However, their approach is bandwidth expensive,
as it requires to exchange an ever growing set of accepted input commands in messages.

In contrast, the protocol presented in \secref{sec:protocol} satisfies a weaker
liveness condition called finite writes termination (FW-termination)~\cite{DBLP:conf/podc/AbrahamCKM04} if
a quorum of processes is correct. Informally, FW-termination guarantees that
write operations (updates) always terminate, whereas read operations (queries) are
guaranteed to terminate only if a finite number of concurrent writes is present. Every FW-terminating protocol
is also lock-free and by extension obstruction-free~\cite{DBLP:conf/icdcs/HerlihyLM03}.

To guarantee the progress of query operations of our protocol, the
number of update operations that occur in parallel with queries must
be limited by a contention management mechanism such as a leader
oracle~\cite{DBLP:conf/podc/AbrahamCKM04}, which is applicable but
beyond the scope of this paper. However, we show in our evaluation
(\secref{sec:evaluation}) that more than 99\,\% of queries can be
processed within one to three round-trips without using such
mechanism.

In the remainder of this section, we sketch an argument for the FW-termination
of our protocol if proposers use incremental prepares to retry failed queries
attempts:

Trivially, all update operations terminate within a single round-trip.
As there are a finite number of updates, there is a point in time in which
the \textit{apply\_update} function is called for the last time, i.e., no
new updates are included in any acceptor. Any proposer that is executing a query
after this point will execute incremental prepares (possibly interleaved with
fixed prepares) until it learns a state. Each time an incremental prepare
is executed, the proposer will either learn a state by consistent quorum or receive at least
one reply with a different payload. If the request fails, the proposer retries with
the LUB of all received payloads from the previous iteration. In each unsuccessful
iteration, the updates of at least one additional acceptor are included in the LUB.
As there is a finite number of acceptors, eventually all acceptors include all updates
and the proposer learns a state by consistent quorum.

\subsection{Optimizations} \label{sec:optimizations}
The base protocol described in \secref{sec:protocol} can be optimized in several ways.
\begin{description}[listparindent=\parindent, topsep=0pt, leftmargin=0cm]
\item[Improve convergence.]
In \algoref{fig:code_crdt_paxos}, proposers include $s_0$ in their initial $\mathit{PREPARE}$
messages. However, computing the LUB with $s_0$ will never increase the payload of an acceptor.
Instead, proposers can either include a payload known from a previous request, or the
payload of a co-located acceptor.

\item[Sending less payloads.] Acceptors do not need to include payloads
in $\mathit{VOTED}$ messages, as this is the state they received from the proposer.
Instead, proposers can simply remember the proposed payload and apply the
queries on it once a quorum of responses is received.

\item[Using delta-mutators.] In the algorithm described above, full CRDT payload
values are sent repeatedly by proposers and acceptors in messages. This can cause high
bandwidth overhead for larger CRDTs. To prevent this, delta-mutators can
be used, as described by Almeida et al.~\cite{almeida2015efficient}. With delta-mutators
it suffices to send state-deltas instead of full values. This can be combined
with techniques introduced by Enes et al.~\cite{enes2019efficient} to
further reduce the amount of redundantly
transmitted data.

\item[Batching.]
Batching is a common strategy to reduce synchronization overhead and bandwidth
needs in workloads with high concurrent access by sacrificing some latency~\cite{DBLP:conf/hpdc/FriedmanR97}.
Implementing batching on a per-proposer basis is simple.
Each proposer manages a separate update and query batch in which it buffers all commands
it has received since the previous batch. To process an update batch, the respective
proposer applies all update commands in the batch on the local replica and
then executes the update protocol normally. For a query batch, the proposer
executes the query protocol and then applies all batched queries on the learned
value.

By the design of our protocol, only a single CRDT payload value is
transferred. It is independent of the size of the batch. Thus, the required
bandwidth only depends on the CRDT's size. In contrast, other
approaches that provide GLA-based state machine replication agree on
command sets, which means that the full batch of commands must be
transmitted (see \secref{sec:related_work}).

\end{description}

\section{Evaluation} \label{sec:evaluation}
\ifanonymous{
  We implemented our protocol in Erlang and tested the implementation's correctness by
  using a protocol scheduler that enforces random interleavings of incoming messages.
}{
  We implemented~\cite{scalaris_approach_impl} our protocol as part of the distributed key-value store Scalaris~\cite{schuett2008scalaris},
  which is written in Erlang. The implementation's correctness was tested using a protocol scheduler that
  enforces random interleavings of incoming messages.
}
For comparison, we use open-source Erlang implementations of
Multi-Paxos~\cite{github_riak_ensemble, Lamport_2001}
and
Raft~\cite{github_raft, Ongaro_Ousterhout_2014}.
We configured both approaches to write their respective command logs
on a RAM disk to minimize their performance impact.  The protocol
proposed by Faleiro et
al.~\cite{Falerio_Rajamani_Rajan_Ramalingam_Vaswani_2012} exchanges an
ever growing set of accepted input commands between its
participants. This set needs to be truncated for this approach to be
practical. Unfortunately, such a mechanism is not described.
As we found that designing one is a non-trivial task, we consider it
out of scope for our evaluation. Thus, the protocol is not included in the
evaluation despite its theoretical importance.

All benchmarks were performed on a cluster equipped with two Intel
Xeon E5-2670 v3 2.4\,GHz per node running Ubuntu 16.04.6 LTS.
The nodes are fully connected with 10\,Gbit/s. For all measurements, we implemented a replicated counter that is
replicated on three nodes using the respective approaches. In our approach, to which we will refer to as CRDT Paxos,
we implemented a G-Counter as described in \secref{sec:crdt_intro}. We applied the
optimizations outlined in \secref{sec:optimizations}, with the exception of delta-mutators.
 For Multi-Paxos and Raft, we used a replicated
integer as the counter. All experiments were executed using Erlang 19.3.
Up to three separate nodes were used to generate load
using the benchmarking tool Basho Bench~\cite{github_basho_bench}.
All measurements ran over a duration of 10 minutes with request data aggregation in 1\,s intervals. For~\figref{fig:throughput_compare}
and~\figref{fig:lat_compare}, we show the median with 99\,\% confidence intervals (CI). The CI is always
within three percent of the reported medians.

\begin{figure*}[t]
    \includegraphics[draft=false,width=1\textwidth]{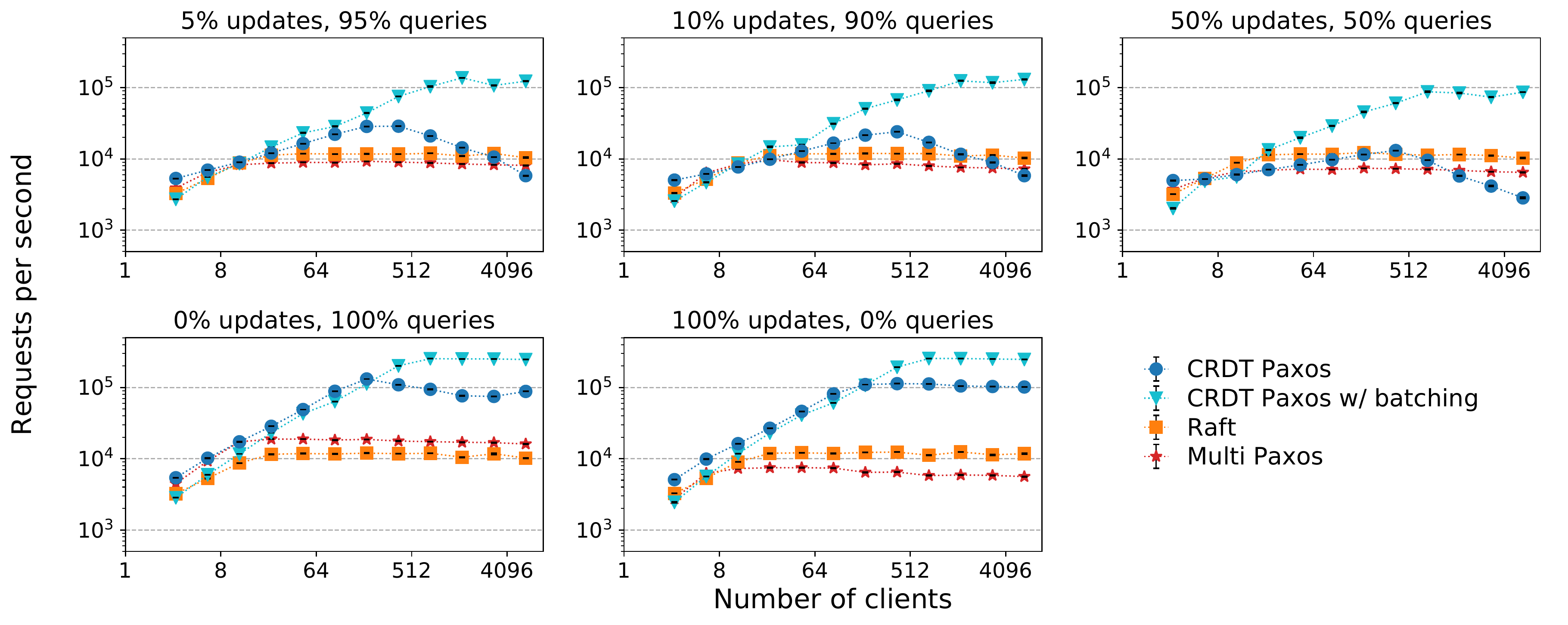}
    \caption{Throughput comparison using three replicas.}
    \label{fig:throughput_compare}
\end{figure*}

\subsection{Failure-free Operation}
In this experiment, we measured the throughput of the approaches under different loads and increasing
number of clients (see \figref{fig:throughput_compare}), which were distributed evenly across three load generators.
Each client independently invokes requests to one of the three replicas and then waits for a response before
invoking the next request. CRDT Paxos performs better for query heavy workloads
as it distinguishes between query and update requests. A decrease in update increases the probability
of observing a consistent quorum, which also increases the ability to process requests in a single round trip.
In contrast, both the Raft and Multi-Paxos implementation append updates and consistent queries to its command log,
which results in their consistent performance for all load types. Overall, CRDT Paxos achieves a higher throughput
for mixed workloads with a low percentage of updates and less than 1500 clients.
This is mainly due to its better load distribution across all replicas compared to the
leader based designs. For more clients, its performance degrades because of the interference between updates and
queries. Note that the $95^{\mathrm{th}}$ percentile query latency of our approach is slightly higher compared to the other
approaches as a small percentage of queries must be retried due to update conflicts (see \figref{fig:lat_compare} and
\ref{fig:rt_compare}).  As updates are always answered in a single round trip, their latencies are
consistently low as long as the nodes and network are not saturated.

The issue of query-update conflicts can be resolved by applying a simple batching scheme
(see \secref{sec:optimizations}): Each proposer processes at most one query and one
update command at a time. During the time a request is processed, all new incoming
commands of the same type (query or update) are batched. Once the ongoing request
is completed the respective batch is submitted.

As this scheme limits the
number of concurrently processed commands, the conflict probability is greatly reduced.
Although no leader is used, more than $99$\% of queries were processed within three round trips for
some measured workloads. Thus, this batching scheme achieves similar throughput in mixed
workloads as CRDT Paxos without batching in query- and update-only workloads, which are
conflict-free and can be processed within a single round-trip.

\subsection{Node Failure}
One drawback to leader-based approaches is their brief unavailability during leader failure
and the added complexity of implementing a leader election algorithm. As our approach does not require
a leader, continuous availability can be achieved as long as a quorum of replicas is reachable.
\figref{fig:crash_lat} shows the impact of a node failure on the $95^{\mathrm{th}}$ percentile latency for 64 clients
and 10\,\% updates.
Latencies increase slightly for the base protocol without batching as all the remaining
replicas must be consistent to reach a consistent quorum. This increases the likelihood of
updates interference. In contrast, a failed replica improves the response latency
when using our batching scheme because the number of concurrently proposed batches
is decreased by one.

\renewcommand{\arraystretch}{0.1}
\newcommand\VRule[1][\arrayrulewidth]{\vrule width #1 depth 0.5pt}
\definecolor{color-sep}{gray}{0.93}
\begin{figure*}
	\captionsetup{justification=centering}
	\begin{tabularx}{\textwidth}{ X !{\color{color-sep}\VRule[3pt]} X !{\color{color-sep}\VRule[3pt]} X }
    \includegraphics[draft=false,clip,trim=10 15 16 15,width=0.3\textwidth]{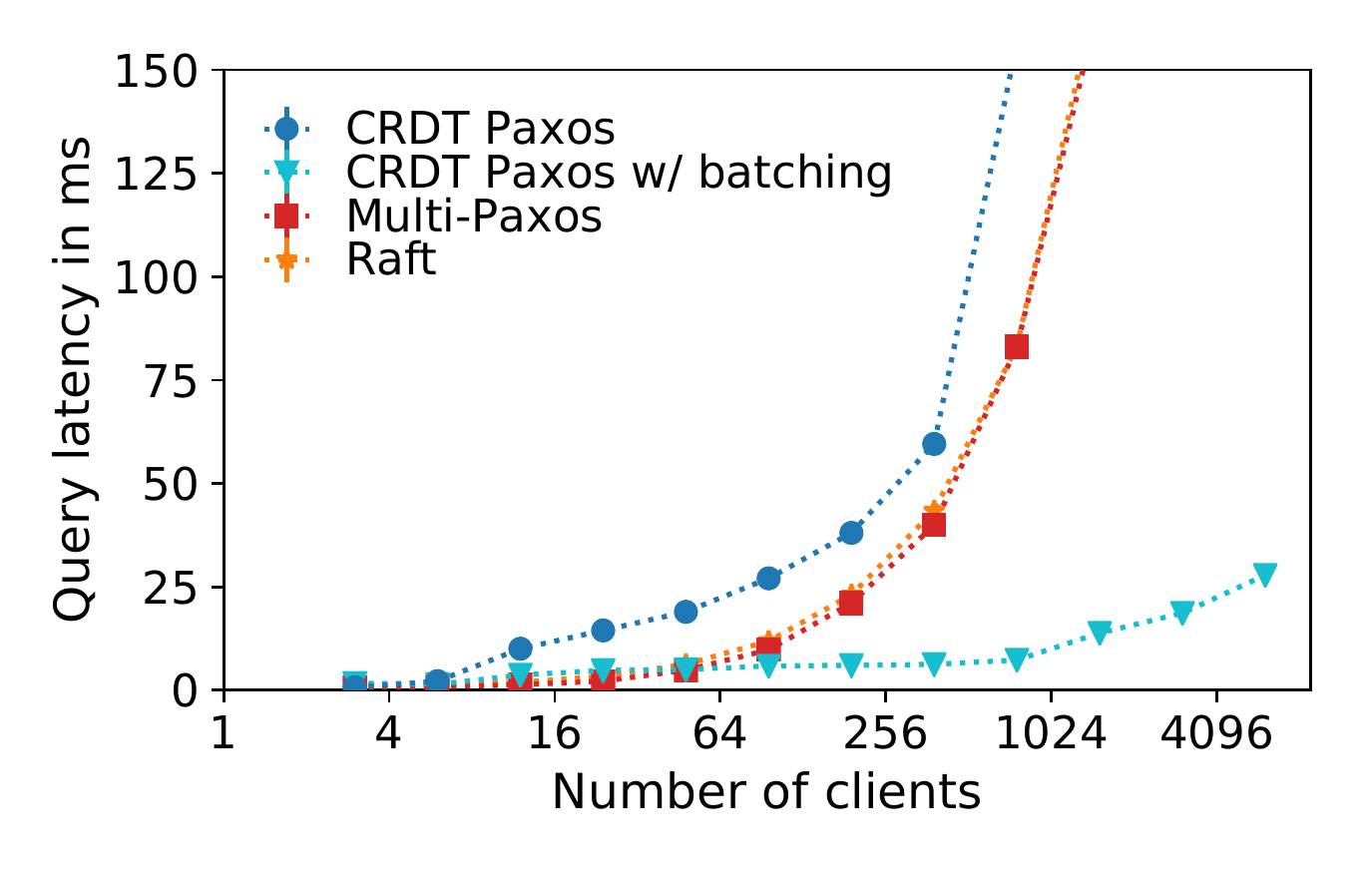} &
    \includegraphics[draft=false,clip,trim=10 15 16 8,width=0.3\textwidth]{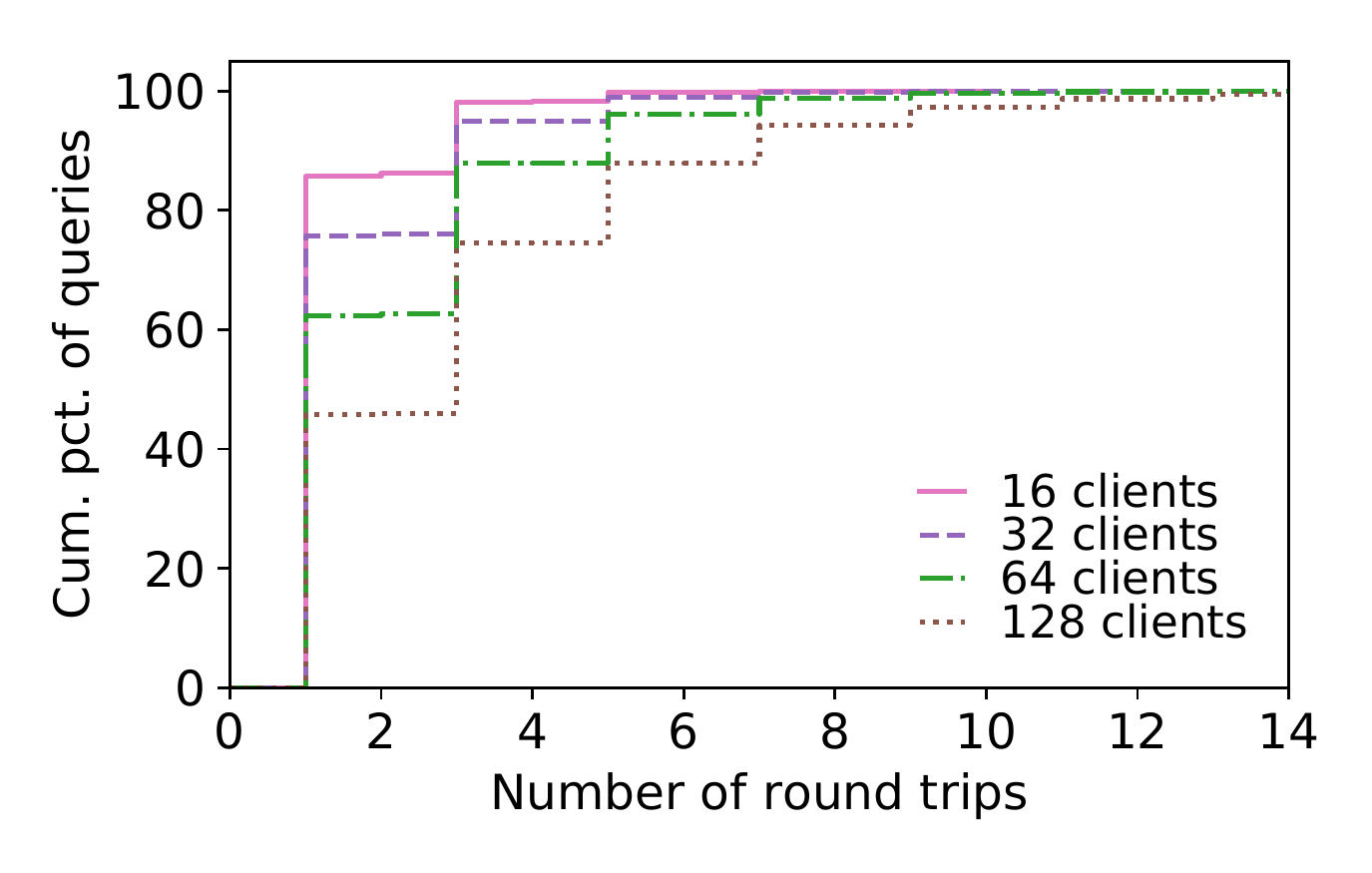} &
    \includegraphics[draft=false,clip,trim=10 15 16 15,width=0.3\textwidth]{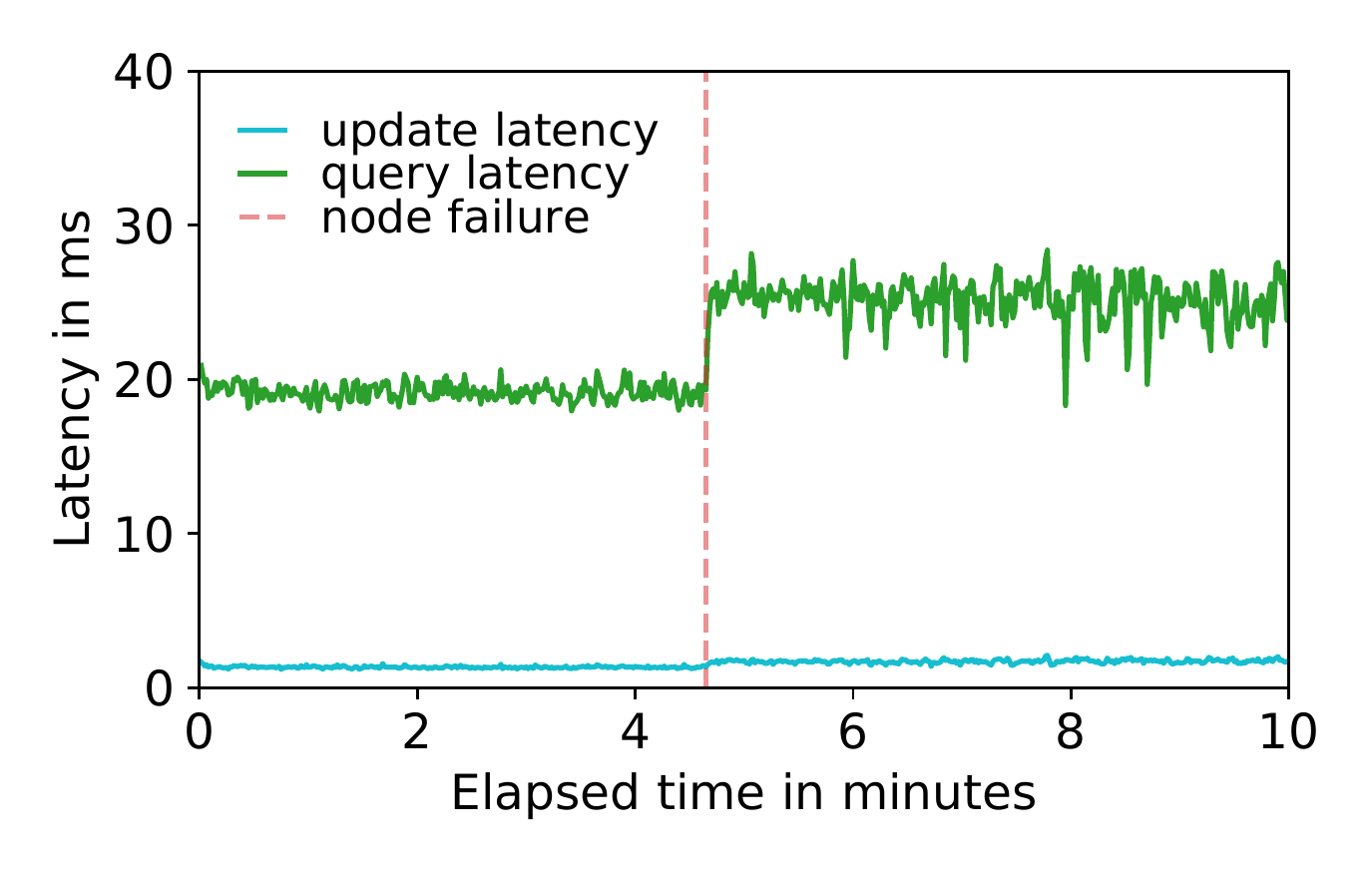} \\

    \begin{subfigure}[t]{0.3\textwidth}
    	\includegraphics[draft=false,clip,trim=10 15 16 15,width=1\textwidth]{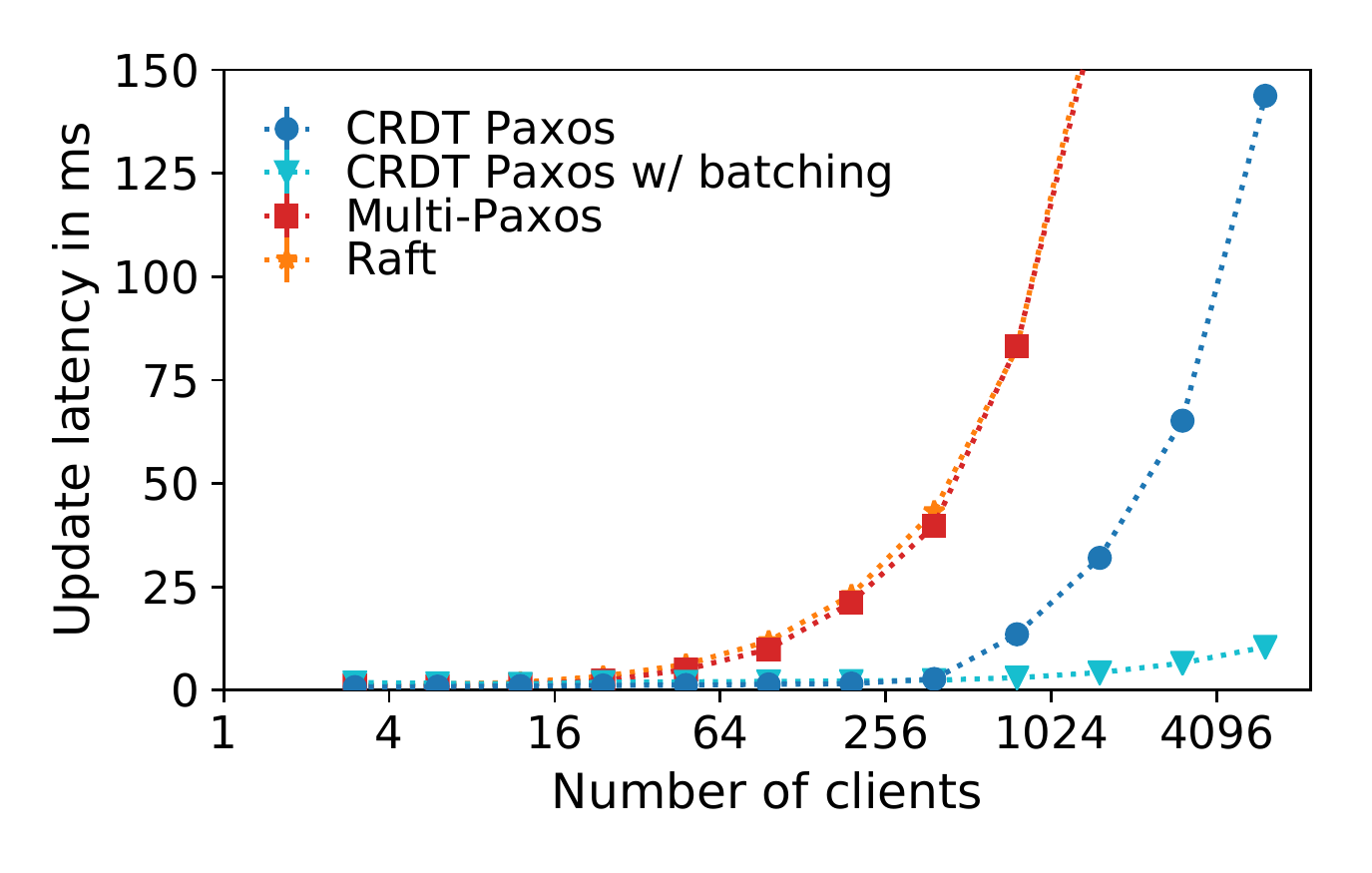}
	    \caption{Query and update latency \\comparison ($95^{\mathrm{th}}$ percentile).}
	    \label{fig:lat_compare}
    \end{subfigure}&
    \begin{subfigure}[t]{0.3\textwidth}
    	\includegraphics[draft=false,clip,trim=10 15 16 15,width=1\textwidth]{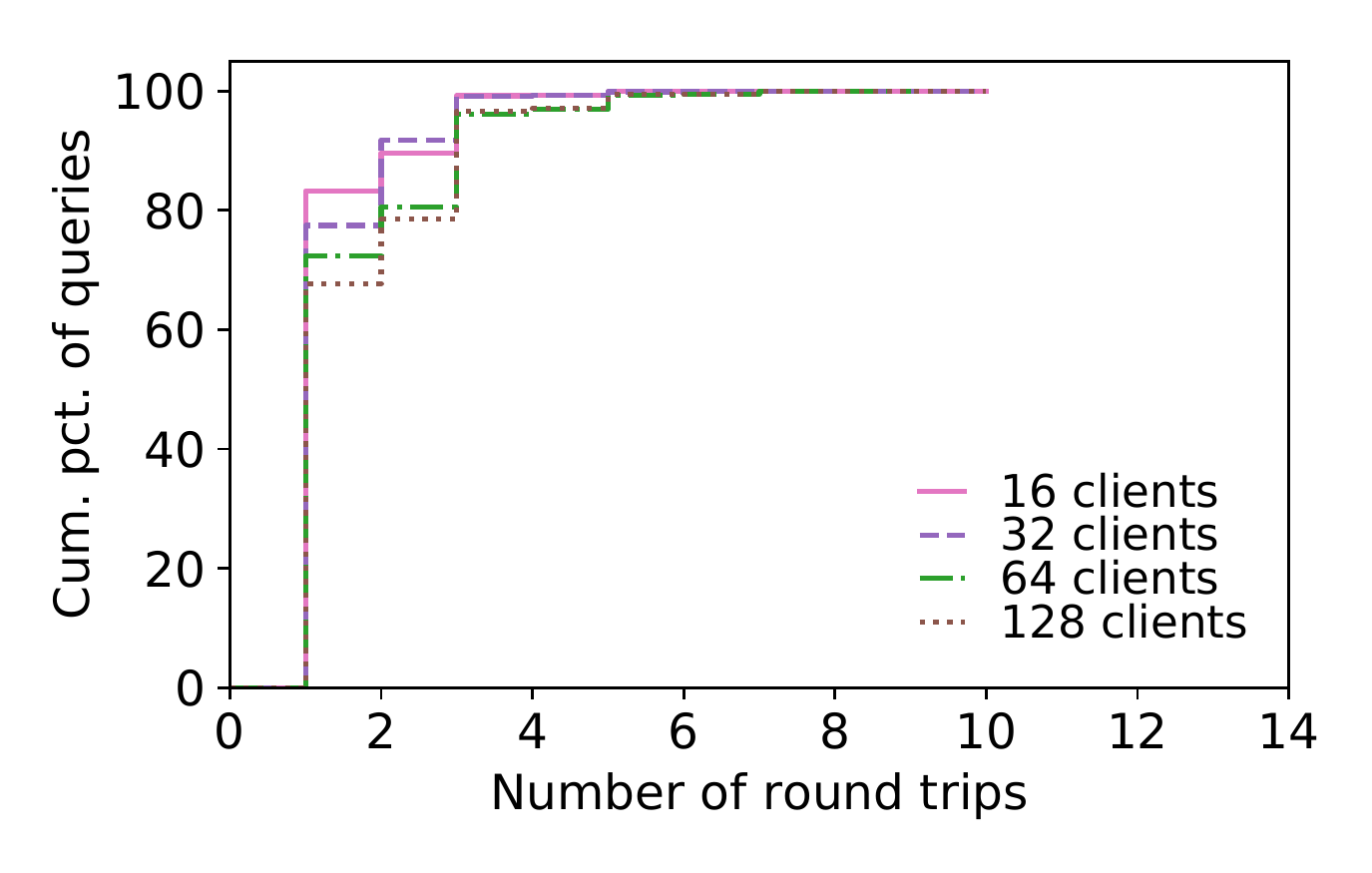}
	    \caption{Round trips to process queries \\(without and with batching).}
	    \label{fig:rt_compare}
    \end{subfigure}&
    \begin{subfigure}[t]{0.3\textwidth}
    	\includegraphics[draft=false,clip,trim=10 15 16 15,width=1\textwidth]{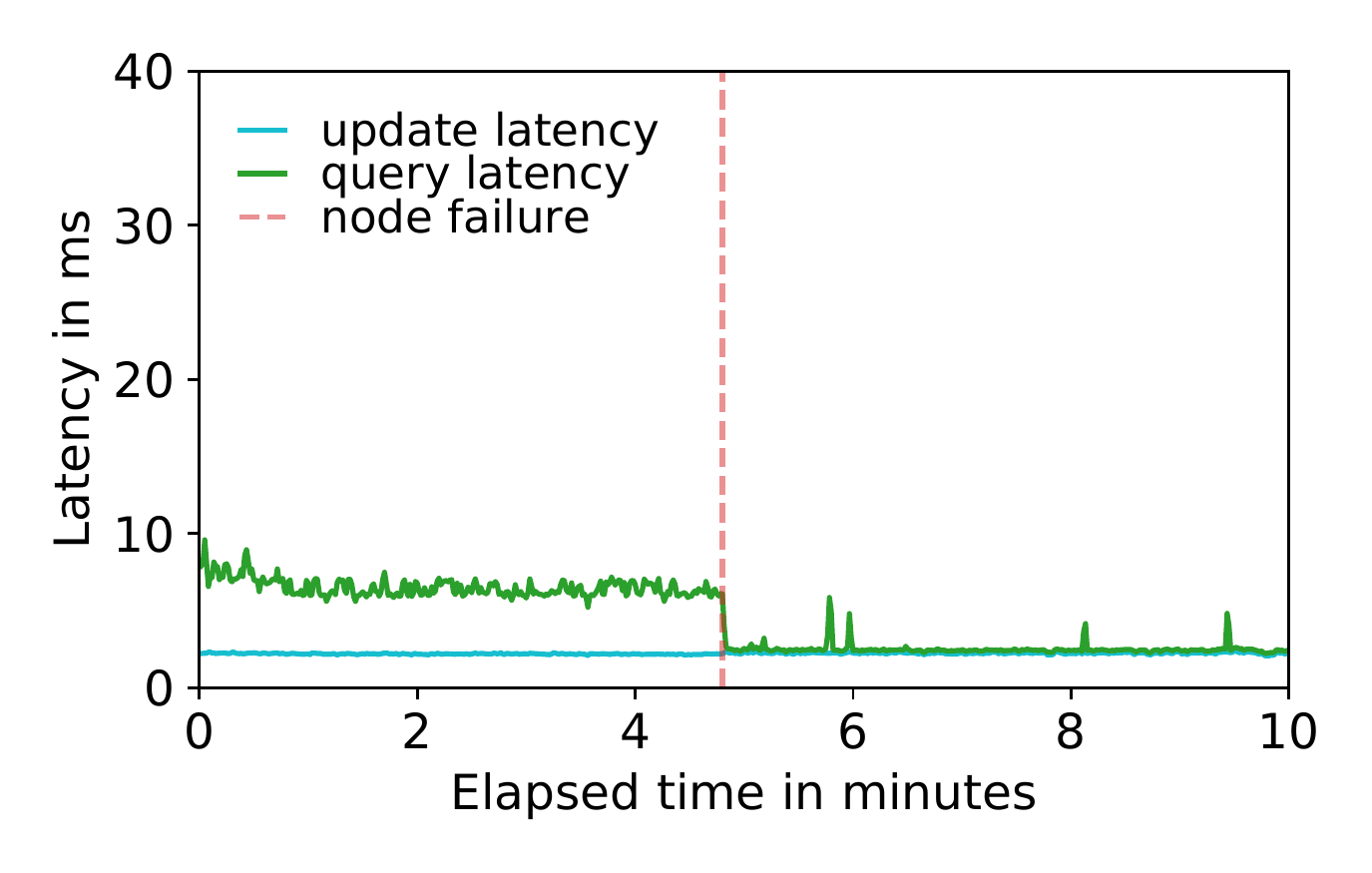}
    	\caption{$95^{\mathrm{th}}$ percentile latency with failure \\(without and with batching).}
	    \label{fig:crash_lat}
    \end{subfigure}\\
	\end{tabularx}
  \caption{Performance evaluation with a query-heavy load (90\% queries) and three replicas.}
\end{figure*}

\section{Related Work} \label{sec:related_work}
As previously mentioned, a wealth of consensus protocols were invented with
the advent of the state machine approach~\cite{schneider1990implementing}, most
notably Paxos~\cite{Lamport_2001,Lamport_1998}, Raft~\cite{Ongaro_Ousterhout_2014}
and variations of them~\cite{Moraru_Andersen_Kaminsky_2013,Hunt_Konar_Junqueira_Reed,Lamport_2006}.
To partially alleviate the high synchronization costs incurred by consensus, numerous
protocols were designed to exploit commutative operations~\cite{Lamport_2005,
Moraru_Andersen_Kaminsky_2013, sutra2011fast}. In contrast to these generalized consensus protocols,
which allow any pair of commands to commute with each other or not, our approach solves
generalized lattice agreement~\cite{Falerio_Rajamani_Rajan_Ramalingam_Vaswani_2012}
by requiring that all update commands commute with each other.
This restriction simplifies the problem so that a high number of concurrent clients
can be supported without the need for a leader or central coordinator. In contrast,
solving (generalized) consensus often relies on efficient leader election~\cite{Ongaro_Ousterhout_2014, aguilera2001stable, malkhi2005omega} or multi-leader approaches~\cite{barcelona2008mencius, camargos2007multicoordinated} to alleviate the
leader performance bottleneck and impact on the system's availability during a
leader failure.

Starting with the original formalization of CRDTs \cite{Shapiro_Preguica_Baquero_Zawirski_2011},
numerous works discuss the design and composition of these data structures~\cite{baquero2016problem, preguica2009commutative, martin2012abstract, bieniusa2012optimized, shapiro2011comprehensive}.
Normal usage of state-based CRDTs require the transmission of the complete
state while dispersing updates to remote replicas. This becomes costly when
CRDTs grow larger. A solution to this problem is discussed by Almeida
et al.~\cite{almeida2015efficient} by only transmitting state-deltas instead of the
complete data structure. Enes et al.~~\cite{enes2019efficient} show how to further
reduce network bandwidth by refining state-delta based synchronization techniques.
Auvolat et al.~\cite{auvolat2019merkle} encode state-based CRDTs into Merkle Search Trees
for efficient access in large networks with high churn and low update rates.

Some CRDT designs suffer from state inflation,
e.g., due to accumulation of tombstone values. Garbage collection mechanisms are
discussed by Shapiro et al.~\cite{shapiro2011comprehensive}. Further
research is needed to find ways to incorporate this into our protocol.

Several protocols that solve generalized lattice agreement in an asynchronous setting exist.
Faleiro et al.~\cite{Falerio_Rajamani_Rajan_Ramalingam_Vaswani_2012} discusses a wait-free protocol
in which a value is always learned in $\mathcal{O}(N)$ messages delays, where $N$ is the number of proposers.
Zheng et al.~\cite{zheng_et_al:LIPIcs:2018:9830} improve this upper
bound to $min\{\mathcal{O}(h(L), \mathcal{O}(f))\}$, where
$h(L)$ denotes the height of the input lattice and $f$ the number of tolerated failures. However, message
sizes can grow unbounded with the number of proposed values in both approaches. Recent work~\cite{zheng2018linearizable}
improves this bound further to $\mathcal{O}(\log{} f)$ round trips and also addresses the problem
of truncating the internally managed command sets. Imbs et al.~\cite{DBLP:conf/icdcn/ImbsMPR18} solves lattice
agreement by introducing a Set-Constrained Delivery (SCD) broadcast primitive, which is build on top of
FIFO broadcast. SCD broadcasting a message requires $\mathcal{O}(N^2)$ messages. All these
approaches agree on growing sets of commands. In contrast, we agree on the resulting CRDT value
directly, which enables some of the optimizations for reducing bandwidth discussed in
\secref{sec:optimizations}.


\section{Conclusion}
In this paper, we presented a protocol that provides linearizable state machine
replication for state-based CRDTs. The protocol guarantees that updates always terminate in
a single round trip. Even though wait-freedom is not provided for query
commands in the presence of concurrent updates,
our experimental evaluation showed that high throughput can be sustained even
under highly concurrent access and without a leader-based
deployment commonly used for consensus-related problems. In addition, our protocol
is lightweight and requires no growing log as it has the memory and message size
overhead of a single counter in addition to the replicated data.
Thereby, no auxiliary processes for
leader election or state management are required for a practical deployment of our approach.
This contrasts our design to the original solution of the generalized lattice
agreement problem~\cite{Falerio_Rajamani_Rajan_Ramalingam_Vaswani_2012}, which is wait-free but requires
additional effort to truncate the managed state or message sizes.

\section{Acknowledgment}
 We thank Alexander Reinefeld for dedicated comments and valuable
 discussions that helped to improve this manuscript. This work was
 supported by the German Research Foundation (DFG) under grant RE 1389
 as part of the DFG priority program SPP 2037 (Scalable Data
 Management for Future Hardware).

\bibliographystyle{IEEEtranS}
\bibliography{main}

\begin{thebibliography}{10}
\providecommand{\url}[1]{#1}
\csname url@samestyle\endcsname
\providecommand{\newblock}{\relax}
\providecommand{\bibinfo}[2]{#2}
\providecommand{\BIBentrySTDinterwordspacing}{\spaceskip=0pt\relax}
\providecommand{\BIBentryALTinterwordstretchfactor}{4}
\providecommand{\BIBentryALTinterwordspacing}{\spaceskip=\fontdimen2\font plus
\BIBentryALTinterwordstretchfactor\fontdimen3\font minus
  \fontdimen4\font\relax}
\providecommand{\BIBforeignlanguage}[2]{{%
\expandafter\ifx\csname l@#1\endcsname\relax
\typeout{** WARNING: IEEEtranS.bst: No hyphenation pattern has been}%
\typeout{** loaded for the language `#1'. Using the pattern for}%
\typeout{** the default language instead.}%
\else
\language=\csname l@#1\endcsname
\fi
#2}}
\providecommand{\BIBdecl}{\relax}
\BIBdecl

\bibitem{DBLP:conf/podc/AbrahamCKM04}
I.~Abraham, G.~V. Chockler, I.~Keidar, and D.~Malkhi, ``{Byzantine Disk Paxos},
  optimal resilience with byzantine shared memory,'' in \emph{{ACM} Symposium
  on Principles of Distributed Computing ({PODC}'04)}, 2004, pp. 226--235.

\bibitem{aguilera2001stable}
M.~K. Aguilera, C.~Delporte{-}Gallet, H.~Fauconnier, and S.~Toueg, ``Stable
  leader election,'' in \emph{Distributed Computing, 15th International
  Conference ({DISC}'01)}, 2001, pp. 108--122.

\bibitem{akka_framework}
{\relax Akka IO}, ``Akka distributed data,''
  \url{https://doc.akka.io/docs/akka/current/typed/distributed-data.html},
  2019, accessed: 2019-05-17.

\bibitem{almeida2015efficient}
P.~S. Almeida, A.~Shoker, and C.~Baquero, ``Efficient state-based {CRDTs} by
  delta-mutation,'' in \emph{Networked Systems - Third International Conference
  ({NETYS}'15)}, 2015, pp. 62--76.

\bibitem{DBLP:journals/jacm/AttiyaBD95}
H.~Attiya, A.~Bar{-}Noy, and D.~Dolev, ``Sharing memory robustly in
  message-passing systems,'' \emph{J. {ACM}}, vol.~42, no.~1, pp. 124--142,
  1995.

\bibitem{auvolat2019merkle}
A.~Auvolat and F.~Ta{\"\i}ani, ``Merkle search trees: Efficient state-based
  crdts in open networks,'' in \emph{SRDS 2019 - 38th IEEE International
  Symposium on Reliable Distributed Systems}, 2019.

\bibitem{baquero2016problem}
C.~Baquero, P.~S. Almeida, and C.~Lerche, ``The problem with embedded {CRDT}
  counters and a solution,'' in \emph{Proceedings of the 2nd Workshop on the
  Principles and Practice of Consistency for Distributed Data,
  (PaPoC@EuroSys'16)}, 2016, pp. 10:1--10:3.

\bibitem{github_basho_bench}
{\relax Basho Technologies}, ``basho-bench: A load-generation and testing tool
  for basically whatever you can write a returning {Erlang} function for,''
  \url{https://github.com/basho/basho\_bench}.

\bibitem{github_riak_ensemble}
------, ``riak\_ensemble: {Multi-Paxos} framework in {Erlang},''
  \url{https://github.com/basho/riak\_ensemble}.

\bibitem{bieniusa2012optimized}
A.~Bieniusa, M.~Zawirski, N.~M. Pregui{\c{c}}a, M.~Shapiro, C.~Baquero,
  V.~Balegas, and S.~Duarte, ``An optimized conflict-free replicated set,''
  \emph{CoRR}, vol. abs/1210.3368, 2012.

\bibitem{soundcloud_roshi}
P.~Bourgon, ``A {CRDT} system for timestamped events,''
  \url{https://developers.soundcloud.com/blog/roshi-a-crdt-system-for-timestamped-events},
  2014, accessed: 2019-08-13.

\bibitem{brown2014riak}
R.~Brown, S.~Cribbs, C.~Meiklejohn, and S.~Elliott, ``Riak {DT} map: a
  composable, convergent replicated dictionary,'' in \emph{Proceedings of the
  First Workshop on the Principles and Practice of Eventual Consistency
  (PaPEC@EuroSys'14)}, 2014, p. 1:1.

\bibitem{camargos2007multicoordinated}
L.~J. Camargos, R.~Schmidt, and F.~Pedone, ``Multicoordinated {P}axos,'' in
  \emph{{ACM} Symposium on Principles of Distributed Computing ({PODC}'07)},
  2007, pp. 316--317.

\bibitem{Chandra_Griesemer_Redstone_2007}
T.~D. Chandra and R.~G.~J. Redstone, ``Paxos made live: an engineering
  perspective,'' in \emph{{ACM} Symposium on Principles of Distributed
  Computing ({PODC}'07)}, 2007, pp. 398--407.

\bibitem{enes2019efficient}
V.~Enes, P.~S. Almeida, C.~Baquero, and J.~Leit{\~a}o, ``Efficient
  synchronization of state-based {CRDTs},'' in \emph{International Conference
  on Data Engineering (ICDE)}, 2019, pp. 148--159.

\bibitem{Falerio_Rajamani_Rajan_Ramalingam_Vaswani_2012}
J.~M. Faleiro, S.~K. Rajamani, K.~Rajan, G.~Ramalingam, and K.~Vaswani,
  ``Generalized lattice agreement,'' in \emph{{ACM} Symposion Principles of
  Distributed Computing ({PODC}'12)}, 2012, pp. 125--134.

\bibitem{Fischer_Lynch_Paterson_1985}
M.~J. Fischer, N.~A. Lynch, and M.~Paterson, ``Impossibility of distributed
  consensus with one faulty process,'' \emph{J. {ACM}}, vol.~32, no.~2, pp.
  374--382, 1985.

\bibitem{DBLP:conf/hpdc/FriedmanR97}
R.~Friedman and R.~van Renesse, ``Packing messages as a tool for boosting the
  performance of total ordering protocols,'' in \emph{Proceedings of the 6th
  International Symposium on High Performance Distributed Computing
  ({HPDC}'97)}, 1997, pp. 233--242.

\bibitem{herlihy1991wait}
M.~Herlihy, ``Wait-free synchronization,'' \emph{{ACM} Trans. Program. Lang.
  Syst.}, vol.~13, pp. 124--149, 1991.

\bibitem{herlihy1990linearizability}
M.~Herlihy and J.~M. Wing, ``Linearizability: {A} correctness condition for
  concurrent objects,'' \emph{{ACM} Trans. Program. Lang. Syst.}, vol.~12,
  no.~3, pp. 463--492, 1990.

\bibitem{DBLP:conf/icdcs/HerlihyLM03}
M.~Herlihy, V.~Luchangco, and M.~Moir, ``Obstruction-free synchronization:
  Double-ended queues as an example,'' in \emph{23rd International Conference
  on Distributed Computing Systems {(ICDCS}'03), 2003}.\hskip 1em plus 0.5em
  minus 0.4em\relax {IEEE} Computer Society, 2003, pp. 522--529.

\bibitem{Hunt_Konar_Junqueira_Reed}
P.~Hunt, M.~Konar, F.~P. Junqueira, and B.~Reed, ``{ZooKeeper}: Wait-free
  coordination for {I}nternet-scale systems,'' in \emph{2010 {USENIX} Annual
  Technical Conference}, 2010.

\bibitem{DBLP:conf/icdcn/ImbsMPR18}
D.~Imbs \emph{et~al.}, ``Set-constrained delivery broadcast: Definition,
  abstraction power, and computability limits,'' in \emph{{ICDCN}'18}, 2018,
  pp. 7:1--7:10.

\bibitem{Lamport_1998}
L.~Lamport, ``The part-time parliament,'' \emph{{ACM} Trans. Comput. Syst.},
  vol.~16, no.~2, pp. 133--169, 1998.

\bibitem{Lamport_2001}
------, ``Paxos made simple,'' \emph{ACM Sigact News}, vol.~32, no.~4, pp.
  18--25, 2001.

\bibitem{Lamport_2005}
------, ``Generalized consensus and {P}axos,'' \emph{Technical Report
  MSR-TR-2005-33}, 2005.

\bibitem{Lamport_2006}
------, ``Fast {P}axos,'' \emph{Distributed Computing}, vol.~19, no.~2, pp.
  79--103, 2006.

\bibitem{malkhi2005omega}
D.~Malkhi, F.~Oprea, and L.~Zhou, ``\emph{Omega} meets {P}axos: Leader election
  and stability without eventual timely links,'' in \emph{Distributed Computing
  ({DISC}'05)}, 2005, pp. 199--213.

\bibitem{barcelona2008mencius}
Y.~Mao, F.~P. Junqueira, and K.~Marzullo, ``Mencius: Building efficient
  replicated state machine for {WANs},'' in \emph{Operating Systems Design and
  Implementation ({OSDI}'08)}, 2008.

\bibitem{martin2012abstract}
S.~Martin, M.~Ahmed{-}Nacer, and P.~Urso, ``Abstract unordered and ordered
  trees {CRDT},'' \emph{CoRR}, vol. abs/1201.1784, 2012.

\bibitem{Moraru_Andersen_Kaminsky_2013}
I.~Moraru, D.~G. Andersen, and M.~Kaminsky, ``There is more consensus in
  {E}galitarian parliaments,'' in \emph{Symposium on Operating Systems
  Principles ({SOSP}'13)}, 2013, pp. 358--372.

\bibitem{Ongaro_Ousterhout_2014}
D.~Ongaro and J.~K. Ousterhout, ``In search of an understandable consensus
  algorithm,'' in \emph{2014 {USENIX} Annual Technical Conference, ({ATC}'14)},
  2014, pp. 305--319.

\bibitem{preguica2009commutative}
N.~Pregui{\c{c}}a, J.~Marqu{\`{e}}s, M.~Shapiro, and M.~Letia, ``A commutative
  replicated data type for cooperative editing,'' in \emph{Distributed
  Computing Systems (ICDCS'09)}, 2009, pp. 395--403.

\bibitem{github_raft}
RabbitMQ, ``ra: A {Raft} implementation for {Erlang} and {Elixir} that strives
  to be efficient and make it easier to use multiple {Raft} clusters in a
  single system,'' \url{https://github.com/rabbitmq/ra}.

\bibitem{redis_crdt}
{\relax Redis Labs}, ``Active-active geo-distribution ({CRDT}-based),''
  \url{https://redislabs.com/redis-enterprise/technology/active-active-geo-distribution/},
  accessed: 2020-07-21.

\bibitem{scalaris_approach_impl}
Scalaris, ``Implementation of \approach/,''
  \url{https://github.com/scalaris-team/scalaris/tree/master/src/crdt}, used
  commit short hash in benchmarks: 8effc6e.

\bibitem{schneider1990implementing}
F.~B. Schneider, ``Implementing fault-tolerant services using the state machine
  approach: {A} tutorial,'' \emph{{ACM} Comput. Surv.}, vol.~22, no.~4, pp.
  299--319, 1990.

\bibitem{schuett2008scalaris}
T.~Sch{\"{u}}tt, F.~Schintke, and A.~Reinefeld, ``Scalaris: reliable
  transactional {P2P} key/value store,'' in \emph{Proceedings of the 7th {ACM}
  {SIGPLAN} workshop on ERLANG}, 2008, pp. 41--48.

\bibitem{shapiro2011comprehensive}
M.~Shapiro, N.~Pregui{\c{c}}a, C.~Baquero, and M.~Zawirski, ``A comprehensive
  study of convergent and commutative replicated data types,'' INRA, Tech.
  Rep., 2011.

\bibitem{Shapiro_Preguica_Baquero_Zawirski_2011}
M.~Shapiro, N.~M. Pregui{\c{c}}a, C.~Baquero, and M.~Zawirski, ``Conflict-free
  replicated data types,'' in \emph{Stabilization, Safety, and Security of
  Distributed Systems ({SSS}'11)}, 2011, pp. 386--400.

\bibitem{crdt-podc-brief}
J.~Skrzypczak, F.~Schintke, and T.~Sch{\"{u}}tt, ``Linearizable state machine
  replication of state-based {CRDTs} without logs,'' in \emph{{ACM} Symposium
  on Principles of Distributed Computing ({PODC}'19)}, 2019, pp. 455--457.

\bibitem{sutra2011fast}
P.~Sutra and M.~Shapiro, ``Fast genuine generalized consensus,'' in \emph{30th
  {IEEE} Symposium on Reliable Distributed Systems (SRDS'11)}, 2011, pp.
  255--264.

\bibitem{viotti2016consistency}
P.~Viotti and M.~Vukoli{\'c}, ``Consistency in non-transactional distributed
  storage systems,'' \emph{{ACM} Comput. Surv.}, vol.~49, no.~1, pp.
  19:1--19:34, 2016.

\bibitem{Vukolic_2012}
M.~Vukoli{\'c}, \emph{Quorum Systems: With Applications to Storage and
  Consensus}, ser. Synthesis Lectures on Distributed Computing Theory.\hskip
  1em plus 0.5em minus 0.4em\relax Morgan {\&} Claypool Publishers, 2012.

\bibitem{zheng2018linearizable}
X.~Zheng, V.~K. Garg, and J.~Kaippallimalil, ``Linearizable replicated state
  machines with lattice agreement,'' \emph{CoRR}, vol. abs/1810.05871, 2018.

\bibitem{zheng_et_al:LIPIcs:2018:9830}
X.~Zheng, C.~Hu, and V.~K. Garg, ``{Lattice Agreement in Message Passing
  Systems},'' in \emph{Distributed Computing (DISC'18)}, vol. 121, 2018, pp.
  41:1--41:17.

\end{thebibliography}

\end{document}